\documentclass[3p]{elsarticle}
\usepackage[utf8]{inputenc}
\usepackage{lineno,hyperref}
\usepackage{amssymb}
\usepackage{amsmath}
\usepackage{xcolor}
\usepackage{epstopdf}
\usepackage{graphicx}
\usepackage{subfig}
\usepackage{multicol}
\modulolinenumbers[5]

\begin{document}

\begin{frontmatter}


\title{Scattered field formalism in the particle-in-cell method for plasma acceleration in tightly focused ultrashort laser beams}

\author[1]{Marianna Lytova\corref{cor1}}
\ead{marianna.lytova@inrs.ca}
\author[2,3]{Fran\c{c}ois~Fillion-Gourdeau}
\author[1,3] {Simon Valli\`eres}
\author[1]{Fran\c{c}ois L\'egar\'e}
\author[2,1,3]{Steve MacLean}

\cortext[cor1]{Corresponding author}

\affiliation[1]{organization=
	\unexpanded{\'Energie, Mat\'eriaux et T\'el\'ecommunications},
	city=\unexpanded{Varennes, Qu\'ebec},
	postcode={J3X 1S2},
	country={Canada}}
\affiliation[2]{organization={Infinite Potential Laboratories},
	city={Waterloo, Ontario},
	postcode={N2L 0A9},
	country={Canada}}
\affiliation[3]{organization={Institute for Quantum Computing, University of Waterloo},
	city={Waterloo, Ontario},
	postcode={N2L~3G1},
	country={Canada}}

\begin{abstract}

The scattered field formalism is combined to the particle-in-cell method to model relativistic laser-plasma dynamics in complex field configurations. Despite the strong nonlinearity of the interactions, we demonstrate the validity of this technique theoretically, by using the semi-linear property of Maxwell-Vlasov equations, and empirically, by performing a systematic convergence analysis with simple 1D and 2D examples. Then, we apply this method to the description of electron acceleration in undercritical plasmas from tightly focused linear and radially polarized beams. For this purpose, an analytical solution for a linearly-polarized tightly focused beam is derived. According to our simulations and their analysis, this approach looks very promising for the optimization of laser-induced relativistic electron beams at ultrahigh intensities because it allows for tightly-focused geometries in reduced calculation volumes. 

\end{abstract}


\end{frontmatter}


\section{Introduction}

For more than 40 years, the acceleration of plasma electrons in a laser field has attracted the attention of researchers because the physical mechanisms underlying this phenomenon are fundamental laser-matter interactions and because it could lead to practical applications. Owing to the high field strengths in infrastructures with intensities above 10$^{22}$ W/cm$^2$ \cite{Bahk:04,Yoon:19}, strong acceleration of electrons on short length scales can be achieved. To take advantage of this, the nonlinear ponderomotive force \cite{Tajima}, laser wake acceleration \cite{Lu, Esarey} and direct longitudinal fields \cite{Gahn, Carbajo, app3010070, 10.1063/1.4738998} were proposed theoretically and realized experimentally. These acceleration techniques allow for relativistic electrons energies above 1 MeV and in the best scenarios, can reach up to the GeV level. To push these results even further, tightly focused configurations have been investigated to exploit the highest intensity range, where quantum electrodynamics effects start to be important \cite{RevModPhys.78.591,PhysRevE.104.015203,PhysRevA.95.032121,bulanov_2022,FEDOTOV20231}. These tightly-focused ultrashort laser pulses are also needed to generate ultrashort relativistic electron beams, which are in demand in medicine and other areas of application \cite{Wen19, vallieres2022}.

To support these investigations, accurate numerical models describing these processes are required to optimize the characteristics of the generated relativistic electron beam, such as the maximum energy of accelerated particles and angular distribution. The self-consistent system of Vlasov-Maxwell equations is one of the most precise theoretical models to describe the dynamics of fields and currents in a collisionless plasma \cite{Lifshitz1981}, but it is computationally intensive. Since the 1960s, particle-in-cell codes (PICs) have been commonly used to solve these equations \cite{Dawson}, being not only computationally efficient, but reasonably accurate if the medium is in the right density range. Modern PIC codes such as EPOCH \cite{Arber_2015}, WarpX \cite{10046112}, PIConGPU \cite{PIConGPU2013} or Smilei \cite{Smilei18} use state-of-the-art parallelization schemes and allow researchers to use this powerful numerical method with minimal coding work by simply setting the laser beam parameters, plasma particle types and simulation window properties. For example, to create a laser beam, one usually needs to specify the boundary on which the laser should originate. In most PIC codes, we can then set any spatiotemporal profile for the magnetic field components at this boundary using built-in or user-defined functions. Unfortunately, this standard approach does not work well with tightly-focused fields, when the beam waist is smaller than the central wavelength and the angular spread is large compared to the paraxial case. Although it is possible to obtain exact or approximate analytical spatiotemporal solutions for tightly-focused laser fields of various types of polarization \cite{April_2010, Salamin:06, Salamin15, Zheng_22}, we cannot properly set them through standard profiles at the boundaries while keeping a reasonable simulation box size. This becomes even more complicated when fields near the focus have to be computed numerically \cite{Jeong:15,Dumont_2017,Vallieres:23,Majorosi:23}. Therefore, alternative numerical methods are needed to insert tightly-focused fields in PIC simulations.

Some techniques have been developed to reach this goal. In \cite{PhysRevE.104.015203}, a short tightly-focused laser pulse is evaluated numerically from a field calculator based on Stratton-Chu integrals. This pulse is then used as an initial condition in PIC simulations, positioned beside the plasma region and thus, requiring large simulation boxes. In Ref. \cite{THIELE20161110}, a different approach is presented where the field is prescribed on a plane in the simulation box and back-propagated to the boundary. In Ref. \cite{Wen19}, an approximate stationary analytic solutions for the non-paraxial case multiplied by some specific time-envelope was implemented in the scattered field formalism (SFF), where the total field is divided into an analytical and a scattered parts \cite{Kunz}. Recently a similar approach, referred to as the analytic pulse technique, was justified in \cite{weichman2023analytic} for implementation with the PIC algorithm. The authors of the latest paper analyze the limitations caused by numerical and physical dispersion and provide simulation examples for relatively long 100 fs pulses containing a large number of cycles. Finally, the SFF has also been used to study particle acceleration in low density gas using longitudinal fields,  \cite{PhysRevLett.111.224801,Marceau_2015}. 

In this paper, we are evaluating the validity of the SFF for ultrashort (here in many examples 2-cycles) tightly-focused fields. The main potential issue when using the SFF for such a configuration is the nonlinear plasma response near the focus as the density approaches the plasma critical density. In this case, the splitting between the incident and scattered field may start to fail, since the linear superposition of solutions breaks down due to the nonlinear nature of the Maxwell-Vlasov equations. For this reason, it is widely believed that the SFF is only accurate for very low densities where nonlinear effects are negligible (as in \cite{PhysRevLett.111.224801,Marceau_2015}). In this article, we argue to the contrary by providing an analytical argument based on the semi-linear property of Maxwell's equation and by performing a systematic numerical convergence analysis. Once the technique is validated in simple configurations, it is applied to the simulation of an experiment where electrons were accelerated in ambient air from tightly-focused fields \cite{vallieres2022} - a case that cannot be properly treated using only the standard PIC code. To simulate the physical conditions in the experiment, we also derive a tightly-focused linearly polarized analytical solution based on April's technique \cite{April_2010}.  

This article is organized as follows. First, in section~\ref{num_tests}, general considerations are given on the compatibility of the SFF with PIC-like codes. These theoretical arguments are supported by simulations performed for 1D and 2D models in parts~\ref{1D} and \ref{2D}, respectively. Section~\ref{3D} presents 3D solutions for electrons acceleration in linearly and radially polarized tightly-focused beams. Comments on performance and general conclusions are made in the final section \ref{concl}.

\section{Scattered field formulation in PIC}\label{num_tests}
The SFF is a standard method in electromagnetics to insert a laser beam in a simulation box \cite{Kunz}. In this case, as a result of Maxwell's equation linearity, it is possible to split the total field in two parts: the incident field, which is a known solution in vacuum, and the scattered field, which contains modifications due to the propagation medium. The SFF presents a number of advantages: first it can accommodate for complex field configurations, second it facilitates the handling of boundary conditions and finally, it can be used to reduce the calculation volume around scattering objects. For these reasons, it has been used extensively for electromagnetic simulations of complex systems \cite{Taflove_2000, Capoglu:13, Wen19, weichman2023analytic}.  

In the PIC method, Maxwell's equation are non-linear because they are coupled to the Vlasov equation, which gives an evolution equation for the phase-space distribution $f_{s}$ for each particle specie $s$ of mass $m_{s}$ and charge $q_{s}$. The actual system of equations (the Maxwell-Vlasov equations) is given by \cite{Lifshitz1981}
\begin{eqnarray}
    \label{eq:vlasov}
    \left(\partial_{t} + \frac{\mathbf{p}}{m_{s} \gamma} \cdot \nabla + q_{s}(\mathbf{E} + \mathbf{v} \times \mathbf{B}) \cdot \nabla_{\mathbf{p}} \right) f_{s}(t, \mathbf{x},\mathbf{p}) &=& 0, \\
	\label{MAmper}
	\partial_t{\bf E} &=& \nabla\times{\bf B} -{\bf J},\\
	\label{MFaraday}
	\partial_t{\bf B} &=& -\nabla\times{\bf E},
\end{eqnarray}
where the relativistic factor is $\gamma = \sqrt{1 + \mathbf{p}^{2} / m_{s}^{2}}$, the velocity is $\mathbf{v} = \mathbf{p}/\gamma m_{s}$,  $\mathbf{E}$ is the electric field, $\mathbf{B}$ is the magnetic field and $\mathbf{J}$ is the current induced by the moving charges, given by
\begin{eqnarray}
    \mathbf{J} = \sum_{s} q_{s} \int d^{3} \mathbf{p} \mathbf{v} f_{s}(t, \mathbf{x},\mathbf{p}),
\end{eqnarray}
that makes the electromagnetic field self-consistent. 

We now discuss how the PIC algorithm works in SFF. As for the usual PIC algorithm, see e.g. \cite{Smilei18}, the distribution function is ``sampled'' by introducing so-called macroparticles, whose dynamics is determined by a ``particle pusher'' \cite{Boris, Vay, Higuera_2017}, a numerical method that solves the Lorentz force equation. The electromagnetic field is solved on a grid via some numerical scheme like the Yee method. Then the current deposition can be done, using Esirkepov's charge-conserving algorithm \cite{ESIRKEPOV}. Since the particle dynamics in the pusher depends on the field values at the position of the corresponding macroparticle, the algorithm must include field interpolation at every time step. However, in the SFF, the total field is assembled from two parts calculated separately on the previous time step, as demonstrated in the following. 

The system of equations \eqref{MAmper}, \eqref{MFaraday} is semilinear and can be written for the total field vector ${\bf F}({\bf x},t) := \big({\bf E}({ \bf x},t), {\bf B}({\bf x},t)\big)^T$ as
\begin{equation}\label{tot_VM}
	\partial_t {\bf F} = \mathbb{A} {\bf F} + {\bf C}({\bf F}),
\end{equation}
where the operator $ \mathbb{A} := \begin{pmatrix}
	0 & 1 \\
	-1 & 0
\end{pmatrix} \nabla\times$ and the ``source'' vector ${\bf C}({\bf F}) := \big({\bf J}({\bf E}, {\bf B}), {\bf 0} \big)^T$.  We assume that the laser (or incident) fields in vacuum  ${\bf E}_{\mathrm{inc}}({\bf x},t)$, ${\bf B}_{\mathrm{inc}}({\bf x},t)$ are known from analytical solutions or numerical calculations in any space point and at any time
 \begin{equation}
 	\partial_t {\bf F}_{\mathrm{inc}} = \mathbb{A} {\bf F}_{\mathrm{inc}},
 \end{equation}
while the fields in an unperturbed homogeneous plasma can be set equal to zero at initial time $t_{\mathrm{in}}$. Then formally applying Duhamel’s formula, we get an integral equation
\begin{equation}\label{Duhamel}
	{\bf F}({\bf x},t)=	{\bf F}_{\mathrm{inc}}({\bf x},t)+\int_{t_{\mathrm{in}}}^t e^{(t-s)\mathbb{A}}\:{\bf C}({\bf F}({\bf x},s))ds.
\end{equation}
We can consider the integral term in \eqref{Duhamel} as the response of the medium to the incident field. This non-linear term describes \textit{scattered} fields generated by currents that have arisen by this time under the action of incident and scattered fields
\begin{equation}\label{scatt_field}
	{\bf F}_{\mathrm{scatt}}({\bf x},t)=	\int_{t_{\mathrm{in}}}^t e^{(t-s)\mathbb{A}}\:{\bf C}({\bf F}_{\mathrm{inc}}({\bf x},s)+{\bf F}_{\mathrm{scatt}}({\bf x},s))ds.
\end{equation}
Eq. \eqref{Duhamel} and \eqref{scatt_field} corresponds to the split between incident and scattered fields ${\bf F}({\bf x},t)=	{\bf F}_{\mathrm{inc}}({\bf x},t) + {\bf F}_{\mathrm{scatt}}({\bf x},t)$, with the resulting evolution equation for the last one given by
\begin{equation}
	\partial_t {\bf F}_{\mathrm{scatt}}({\bf x},t) = \mathbb{A} {\bf F}_{\mathrm{scatt}}({\bf x},t) + {\bf C}({\bf F}).
\end{equation}
To close the time cycle of the algorithm, this equation is solved by a numerical PIC scheme on a staggered grid using the FTDT method \cite{Yee} (although other numerical schemes could be used). In the next time cycle, the scattered field calculated in this way is interpolated and added to the incident field at the position of the macroparticle, so both fields participate in the pushing of macroparticles and therefore affect the current deposition. 

Technically, we implemented this algorithm in Smilei using a block called {\ttfamily PrescribedField}, that allows superimposing any user-defined spatiotemporal electromagnetic fields on fields computed using the Maxwell solver. Other PIC codes have a similar functionality. Using {\ttfamily ctypes}, a well-known Python library, allows us to implement any desired field configuration using optimized C-functions, rather than directly including these expressions in Smilei's Python input file, resulting in higher performance.

To summarize, we have demonstrated theoretically that because the system of equations is semi-linear, the SFF can be applied to PIC calculations. In the next section, we present some test examples of solutions obtained in SFF and analyze their convergence.

\section{Convergence analysis for test solutions obtained in SFF}\label{ConvAn}

In this section, some simple 1D and 2D field configurations are used to test the validity of the SFF by performing a systematic convergence studies and by comparing the SFF with the standard method. The general strategy consists of evaluating the same solutions for the interaction of a laser pulse with a plasma but using different implementations: (i) using a laser beam creator at the boundary (standard for PIC codes method) and (ii) using analytical expressions for the components of the incident field in the SFF.

Unless otherwise stated, dimensionless variables are used, which are typical for plasma physics. They are defined in Table \ref{tab:units} in terms of the central laser wavelength $\lambda_c$. 

\begin{table}[h]
\begin{center}
\begin{tabular}{ l c } 
 \hline 
 Unit quantity & Value (SI units) \\
 \hline 
 Charge ($Q_{r}$) & Elementary charge $e$ \\
 Mass ($M_{r}$) & Electron mass $m_{e}$ \\
 Velocity ($V_{r}$) & Speed of light $c$ \\
 Length ($L_{r}$) & $\lambda_c/2\pi$ \\
 Frequency ($\omega_{r}$) & $2\pi c/\lambda_c$ \\
 Time ($T_{r}$) & $1/\omega_r$ \\
 Electric fields ($E_r$) & $m_ec^2/eL_r$ \\  
 Magnetic fields $B_r$ & $m_ec/eL_r$  \\
 Plasma density ($N_r$) & $\epsilon_0 m_e \omega_r^2/e^2$ \\
 Current ($J_r$) & $eN_rc$ \\
 \hline
\end{tabular}
\caption{Dimensionless plasma units \label{tab:units}}
\end{center}
\end{table}

\subsection{1D simulations}\label{1D}

The first numerical test is for a simple Gaussian laser pulse in 1D interacting with a plasma. In case (i) we set a laser beam with normalized central frequency $\omega=1$ (below we  omit it in the notation as well as the normalized wavenumber $k=1$) on the left side of the computational domain $z\in[0, 2\pi\cdot64]$:
\begin{eqnarray}
        \label{Bx1d}
	B_x(z=0,t) &=& 0,\\
        \label{By1d}
	B_y(z=0,t)&=&B_0\sin(t-t_0 - z_{\mathrm{ghost}})\cdot2^{-\frac{4(t-t_0-z_{\mathrm{ghost}})^2}{\text{FWHM}_T^2}}.
\end{eqnarray}
In our simulations, the position of the maximum of the envelope is $t_0=25.6\pi$, the pulse duration is FWHM$_T$=2$\pi$ - ``full width at half maximum'', and the pulse amplitude is $B_0 = 5$. Here we also introduced the parameter $z_{\mathrm{ghost}}$ to compensate for the shift caused by the presence of ghost cells required for the parallelization algorithm. In Smiley's code, this parameter is equal to the interpolation order times the cell size $\Delta z$, whereas the default interpolation order is 2. In this case, if we want the laser beam to originate exactly at the boundary $z=0$, we need use $z_{\mathrm{ghost}}=-2\Delta z$. In the absence of a medium, that is, in a vacuum, starting from the boundary conditions \ref{Bx1d}, \ref{By1d} only the Maxwell solver of the PIC code provides a numerical solution for all field components as functions of $z$, at any time. This numerical solution must match to model (ii), which gives exact analytical expressions for the electric and magnetic components in vacuum:
\begin{eqnarray}
	E_x(z,t)&=&E_0\sin(t-t_0-z)\cdot2^{-\frac{4(t-t_0-z)^2}{\text{FWHM}_T^2}},\\
	B_y(z,t)&=&B_0\sin(t-t_0-z)\cdot2^{-\frac{4(t-t_0-z)^2}{\text{FWHM}_T^2}},
\end{eqnarray}
with the same parameters as in case (i) and $E_0=B_0$. 

\begin{figure}[b!]
	\centering{
		\includegraphics[scale=0.85,trim={0 0 0 0},clip]{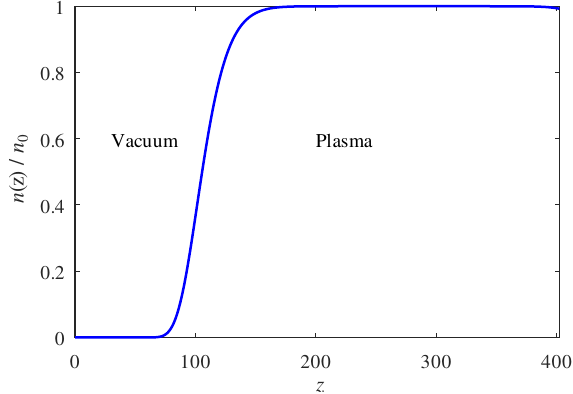}}
	\caption{Initial plasma profile in the simulation region} \label{super_gauss}
\end{figure}

After matching the models in vacuum, we can add plasma to them. In our simulations, a smooth plasma profile that reduces reflection is given by a super-Gaussian distribution with order $p = 12$, centered at $z_0=281.5$, and FWHM$_Z$=335.1, see Fig.~\ref{super_gauss}
\begin{equation}
	n(z)=\left\{
	\begin{matrix}
		0, & z \leq 20\pi\\
		n_0\cdot2^{-\big[\frac{2(z-z_0)}{\text{FWHM}_Z}\big]^p}, & z > 20\pi.
	\end{matrix}
	\right.
\end{equation}
For electrons, $n_{e0}$ is chosen twice as much as for doubly ionized atomic nitrogen $n_{N0}$ - here we assume that preionization takes place. In 1D simulations, we explore two values of the maximum electron density. The first is far from the critical density $n_{c}$ and is set to $n_{e0}/n_{c}=0.1$, while the second is $n_{e0}/n_{c}=0.75$. In this set of simulations, we do not consider either field or collisional ionization during laser-plasma interactions.
We follow the pulse propagation through the computational domain during the simulation time $T_{\mathrm{sim}}=2\pi\cdot64$ and store data for several intermediate points in time $t_n$, $n=1,\dots,N$ with maximum $N=16$. To perform a convergence analysis, we ran simulations for six spatial steps $\Delta z = \dfrac{2\pi}{16\cdot2^m}$, $m=\{0,1,\dots,5\}$, and the corresponding time steps for the 1D case were chosen according to the Courant–Friedrichs–Lewy (CFL) stability condition \cite{Strik,gross} as $\Delta t=0.99\Delta z$. 

\begin{figure}[p]
	\centering{
		\includegraphics[scale=0.5,trim={60 40 60 50},clip]{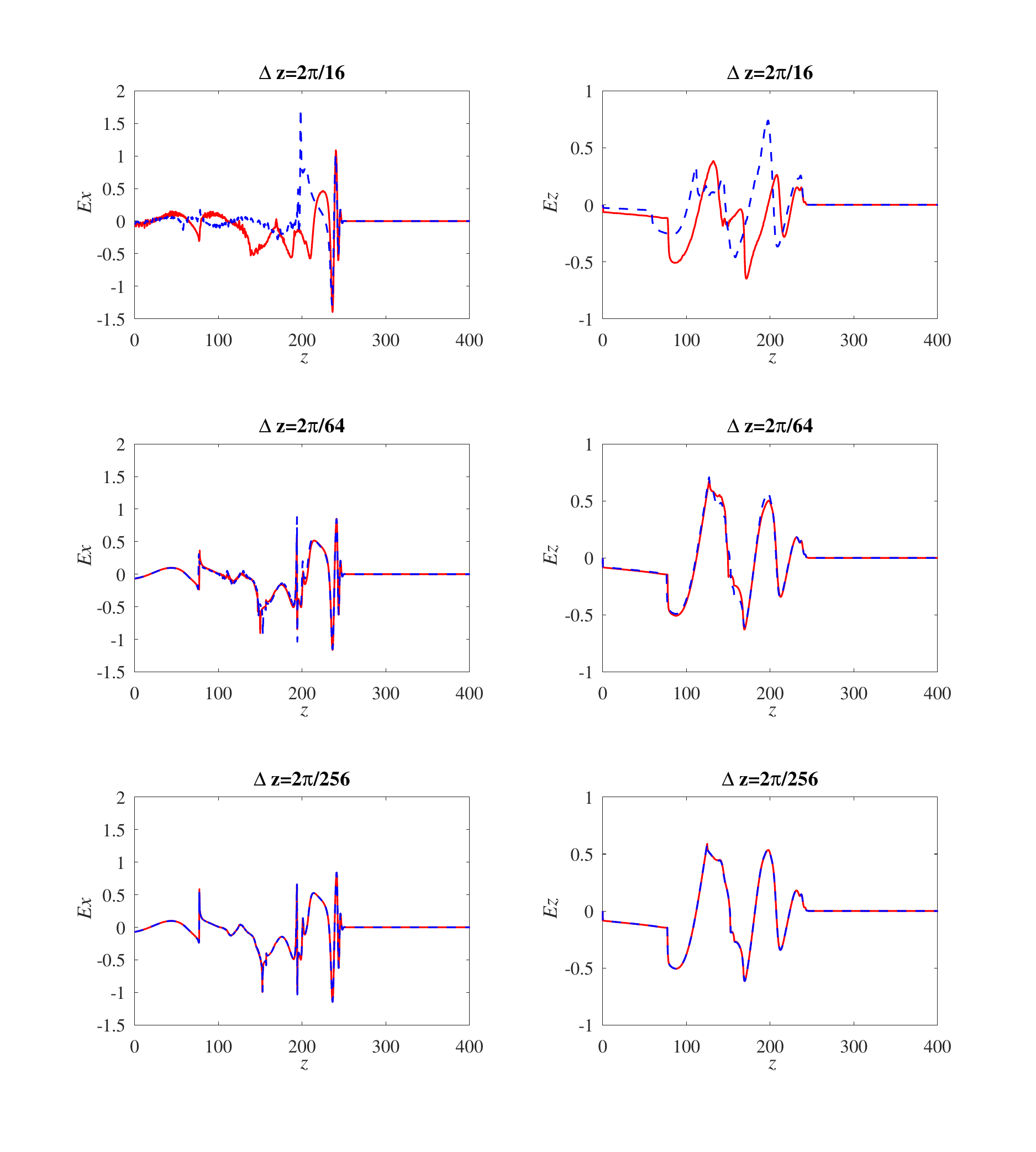}}
	\caption{Electric field components $E_x(z,t)$ and $E_z(z,t)$ calculated with sequential decrease in cell size $\Delta z = \{2\pi/16, 2\pi/64, 2\pi/256\}$ for the problem of laser envelope propagation in plasma with density $n_{e0}=0.1$ at simulation time $t=323$. Solutions shown in blue (dashed lines) are obtained for case (i) when laser field initially was set through the laser creator, and solutions in red correspond to the case (ii) when the incident field is prescribed.} \label{Esol_N01}
\end{figure}

\begin{figure}[p]
	\centering{
		\includegraphics[scale=0.5,trim={60 40 60 50},clip]{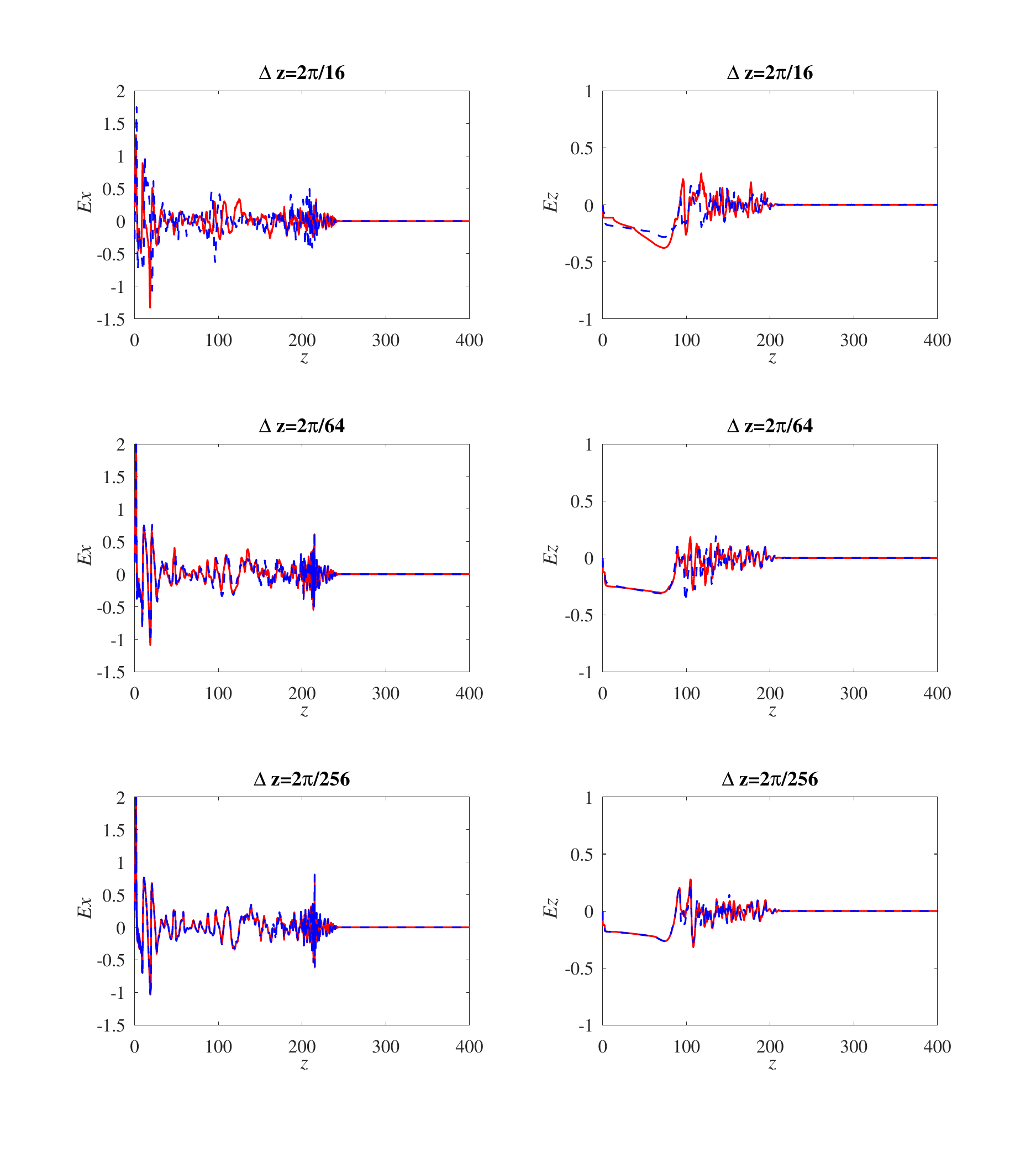}}
	\caption{Electric field components $E_x(z,t)$ and $E_z(z,t)$ calculated with sequential decrease in cell size $\Delta z = \{2\pi/16, 2\pi/64, 2\pi/256\}$ for the problem of laser envelope propagation in plasma with density $n_{e0}=0.75$ at simulation time $t=323$. Solutions shown in blue (dashed lines) are obtained for case (i) when laser field initially was set through the laser creator, and solutions in red correspond to the case (ii) when the incident field is prescribed.} \label{Esol_N075}
\end{figure}

Figures~\ref{Esol_N01} and \ref{Esol_N075}, which correspond to plasma densities of 0.1 and 0.75, respectively, show examples of solutions for two components of the electric field: the transverse component $E_x(z,t)$ modified by the plasma, and the longitudinal component $E_z(z,t)$ arising in the plasma. A somewhat stronger damping of the field amplitude is observed in the case of a denser plasma, accompanied, however, by higher frequencies of field disturbances. For both plasma densities, convergence of solutions is observed with decreasing cell size $\Delta z$. 


The following Figures~\ref{L2_bothN} (a)-(d) demonstrate $L_2$-stability of the numerical solutions $E^{\Delta z}(z_i,t_n)$ saved at times $t_n$ using the discrete norms for errors \cite{Strik}
\begin{equation}\label{L2}
	\|\delta E(\cdot, t_n)\|_{2,\Delta z}=\Big(\Delta z \sum_k |E^{\Delta z}(z_k,t_n)-E^{\mathrm{exact}}(z_k,t_n)|^2  \Big)^{1/2},
\end{equation}
where $E^{\Delta z}=E_{\{x,z\}}^{\{(i),(ii)\}}$ obtained for some $\Delta z$, and solution for cell size $\Delta z=2\pi/256$ was chosen as ``\textit{exact}'', since twice lower step $2\pi/512$ does not noticeably change the results. Note that when calculating most norms, the discrete solutions $E_x$ and $E_z$ refer to the same way of implementing the laser beam: (i) or (ii), but we also show one \textit{cross-norm} for the residue between \textit{exact} solutions obtained in different models:
\begin{equation}\label{L2_cross}
	\|\delta E^{\mathrm{cross}}(\cdot, t_n)\|_{2}=\Big(\frac{2\pi}{256} \sum_k |E^{(i), \mathrm{exact}}(z_k,t_n)-E^{(ii), \mathrm{exact}}(z_k,t_n)|^2  \Big)^{1/2}.
\end{equation}
\begin{figure}[p]
	\centering{
		\subfloat{\includegraphics[scale=0.5,trim={45 0 60 20},clip]{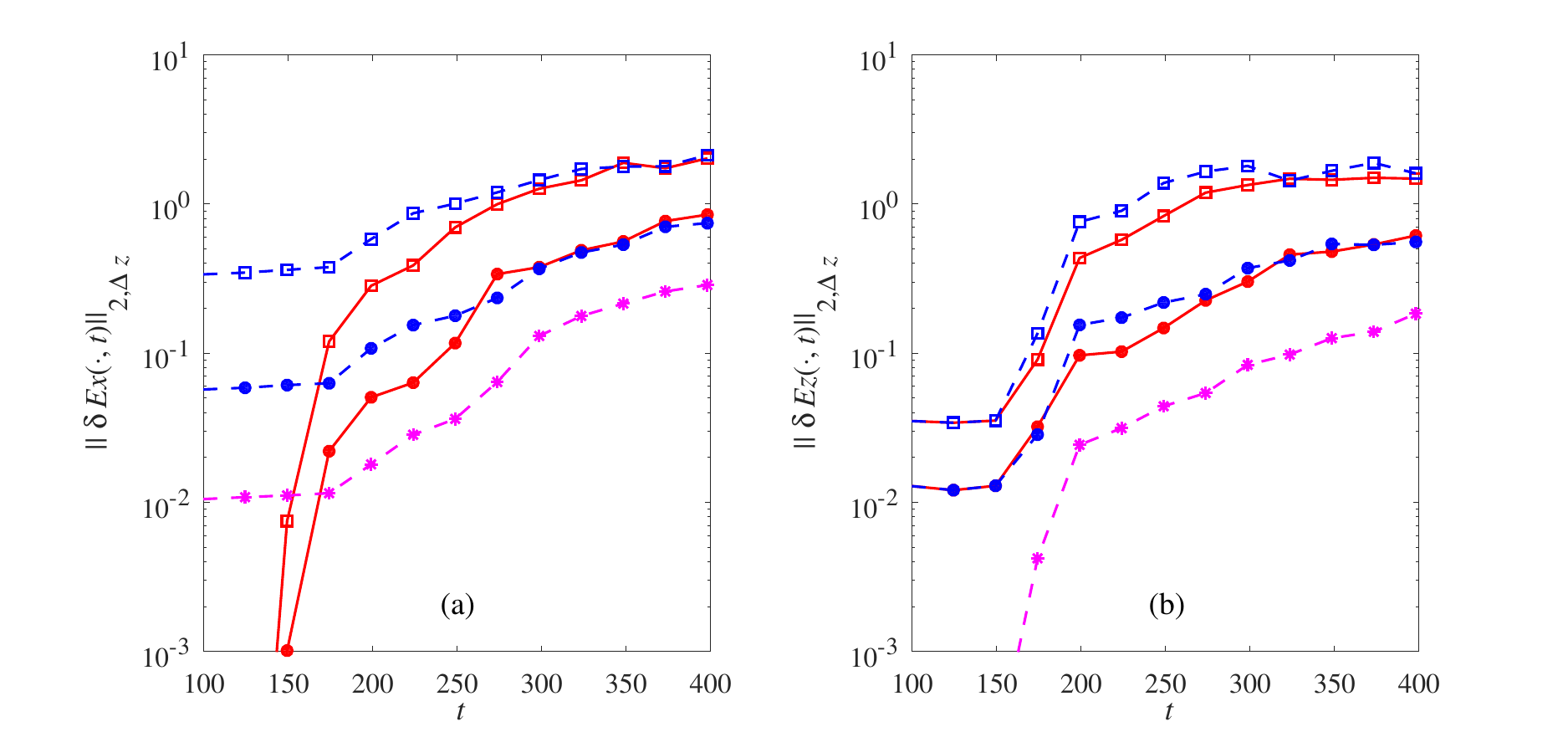}}\\
		\vspace{2mm}
		\subfloat{\includegraphics[scale=0.5,trim={45 0 60 20},clip]{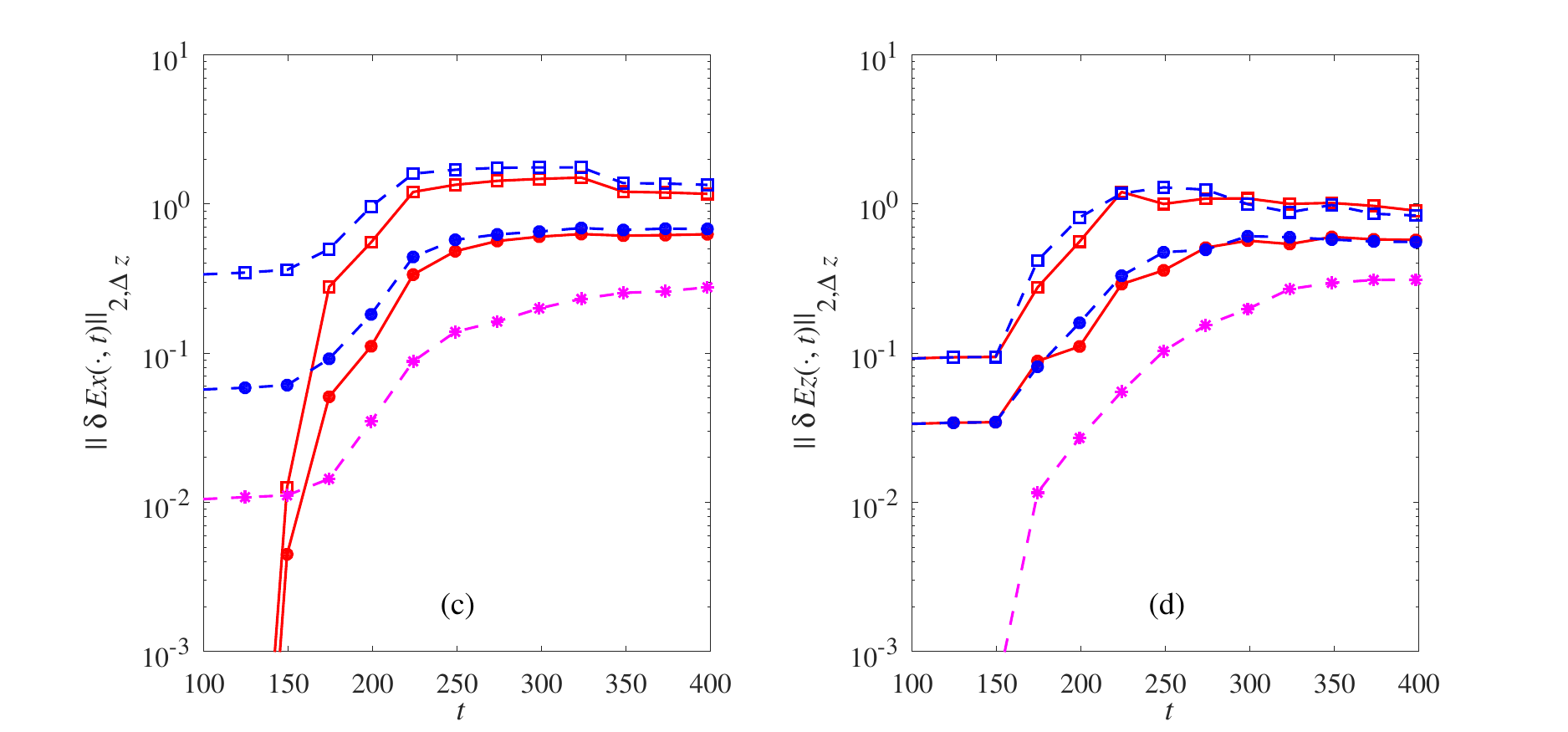}}
	}
	\caption{Discrete $L_2$ norms for errors of solutions $E_x(z_i,t_n)$ - (a),(c) and $E_z(z_i,t_n)$ - (b),(d) computed according to \eqref{L2} for plasma number density $n_{e0}=0.1$ - (a),(b) and $n_{e0}=0.75$ - (c),(d). Open squares are used for $\Delta z= 2\pi/16$, and filled circles - for $\Delta z= 2\pi/64$. The blue curves correspond to the case (i) - laser beam creator, the red curve corresponds to the applying of a prescribed field - case (ii). The norms, shown in magenta with asterisk markers, are calculated through the difference between the solutions for cases (i) and (ii) at $\Delta z=2\pi/256$.} \label{L2_bothN}
\end{figure}
We see that for each plasma density, the same three norms origin from zero (or minus infinity in the $\log$-scale) for  times preceding the entry of the pulse into the plasma. For the $E_x$-component, this happens for solutions obtained in the SFF (ii) (for any $\Delta z$). This is reasonable, since in vacuum, we use analytical solutions that coincide at common grid points, and in the 1D model there is no non-zero $E_x$ portion, which follows from the solution of Poisson's equation. (The latter is included in the PIC code to calculate the initial electric field of the plasma particles) On the contrary, the solutions $E_x$ obtained for case (i), i.e. from the condition on the left boundary, are fully numerical solutions for all times and their norms increase, albeit very slowly, when propagating in vacuum. In the case of the longitudinal component $E_z$, the norms behave differently, since initially this component is calculated by the Poisson's equation in the unperturbed plasma and depends only on the discretization step. Thus, for propagation in vacuum, the solutions for this component coincide for models (i) and (ii) (moreover, in vacuum in the limit $\Delta z \rightarrow 0$, $E_z \rightarrow 0$), so that the $\log$ of the cross-norm (curve with asterisks) tends to $-\infty$ as time goes to 0. Therefore, it can be concluded from figures~\ref{L2_bothN} (a)-(d) that in both cases (i) and (ii), the numerical solutions are stable.

In all 1D simulations, we used a number of macroparticles per cell equal to 10,000 for electrons and 5,000 for Nitrogen ions. This number seems to be large enough even for $\Delta z = 2\pi/16$, since using 40,000 macroparticles per cell changes the norm of $\|\delta E_x(\cdot, t_n)\|_{2,\Delta z}$ no more than $1\%$ for all times.

\begin{figure}[ph!]
	\centering{
		\subfloat{\includegraphics[scale=0.5,trim={60 0 60 20},clip]{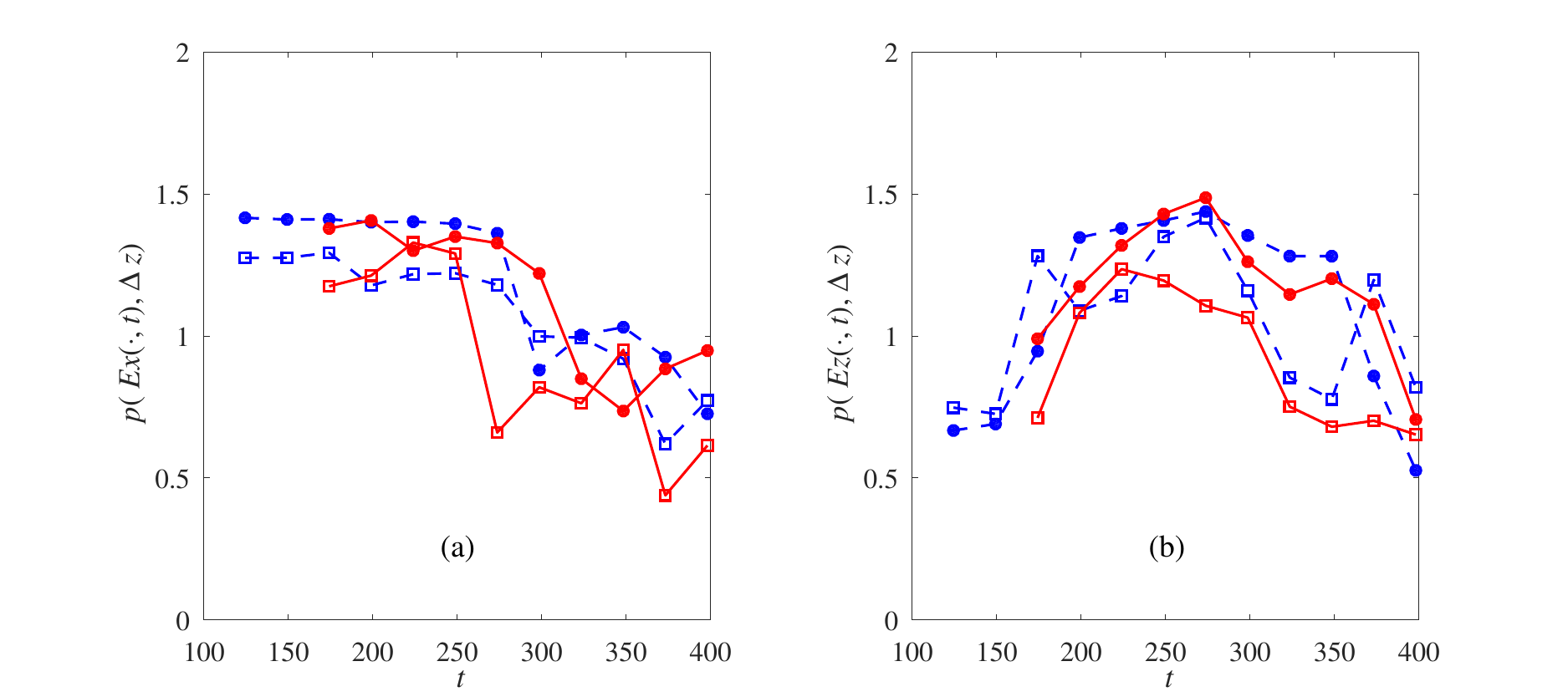}}\\
		\vspace{2mm}
		\subfloat{\includegraphics[scale=0.5,trim={60 0 60 20},clip]{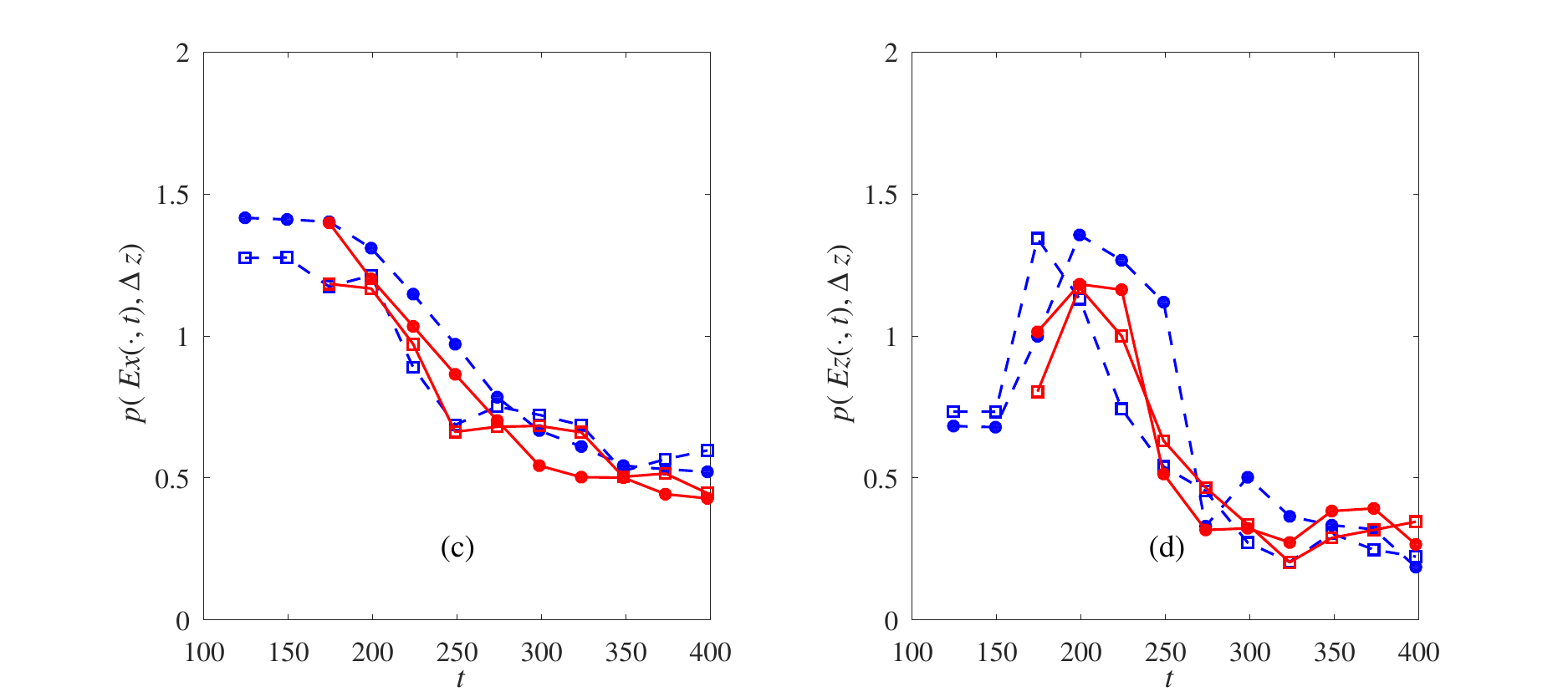}}
	}
	\caption{Numerical convergence rates for solutions $E_x(z_i,t_n)$ - (a),(c) and $E_z(z_i,t_n)$ - (b),(d) computed according to \eqref{p_rate} for plasma number density $n_{e0}=0.1$ - (a),(b) and $n_{e0}=0.75$ - (c),(d). Open squares are used for $\Delta z= 2\pi/16$, and filled circles - for $\Delta z= 2\pi/64$. The blue dashed curves correspond to the case (i) using a laser creator, the red curve corresponds to the applying of a prescribed field - case (ii).} \label{Cnv_bothN}
\end{figure}

It is well known that the order of consistency for Yee's scheme is 2 \cite{Yee}. Next, we present our calculations of the numerical rates of convergence. Based on the consistency of discretization method and its stability we can write 
\begin{equation}
	\|\delta E(\cdot, t_n)\|_{2,\Delta z} \leq \mathcal{C}(\Delta z)^p,\quad \mathcal{C} - \text{const.}
\end{equation}
whence we get the following measure 
\begin{equation}\label{p_rate}
	p(E(\cdot, t_n), \Delta z)=\log_2\Big(\dfrac{\|\delta E(\cdot, t_n)\|_{2,\Delta z}}{\|\delta E(\cdot, t_n)\|_{2,\Delta z/2}}\Big).
\end{equation}
Note that although stability of $L_2$-norm imply stability of $L_1$-norm as well, in \eqref{p_rate} we use the $L_2$ discrete norm. Usually this is  a lower bound on the convergence rate computed through $L_1$-norms. The convergence rates calculated in this way are shown in Figures~\ref{Cnv_bothN}~(a)-(d). Despite the unevenness of these parameters, we see that both models (i) and (ii) show close orders of convergence over the entire propagation time. Due to the stronger order of non-linearity in plasma with $n_{e0}=0.75$, we see faster decreasing in values of $p$ over time, however, this is true for both models.

\begin{figure}[ht!]
	\centering{
		\includegraphics[scale=0.5,trim={60 40 60 50},clip]{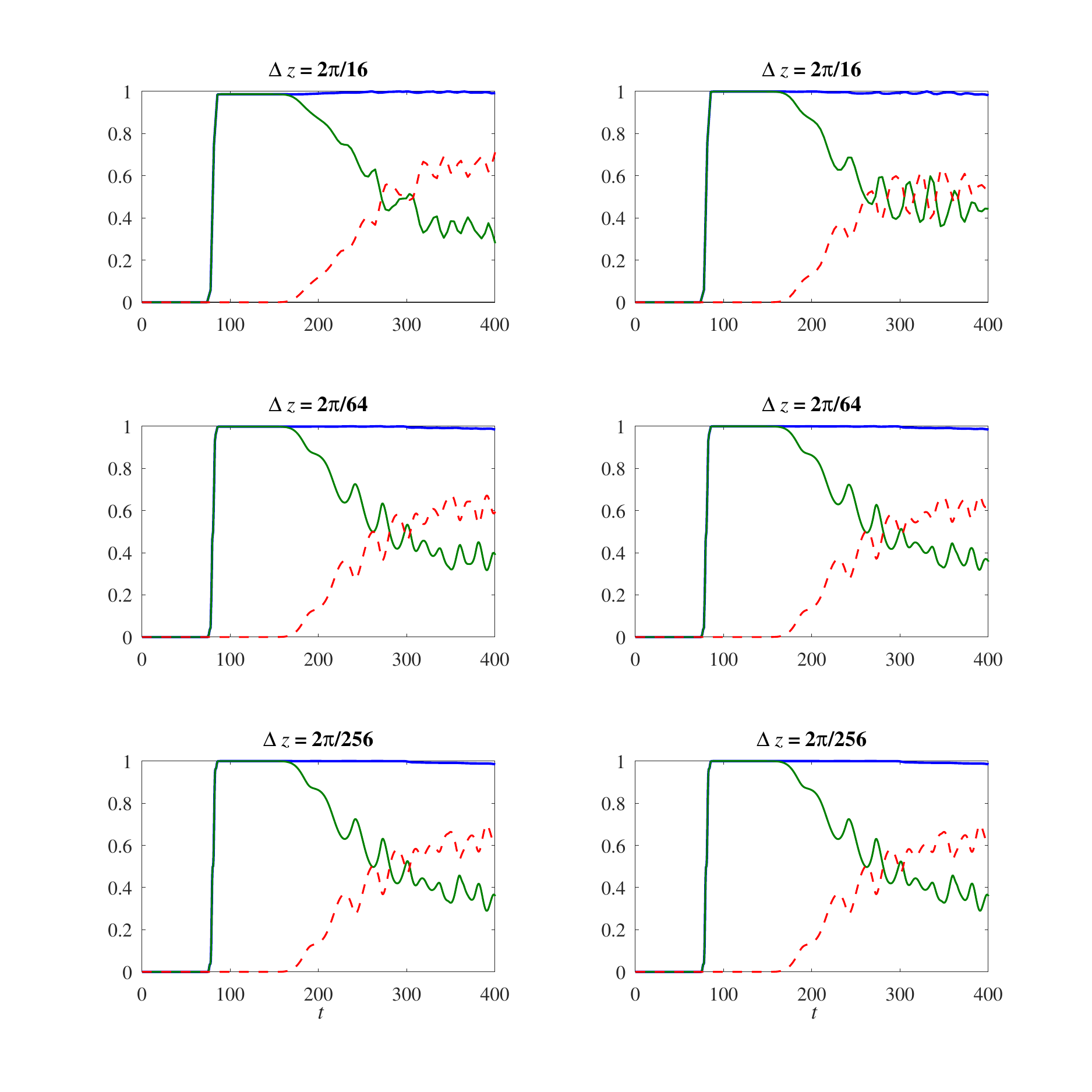}}
	\caption{The integrated kinetic energy of electrons (dashed red curves), the integrated energy of the electromagnetic field (thin green curves), the total energy (thick blue curves) in time, calculated for a pulse propagating in a plasma with a density of $n_{e0}=0.1$ in models with a laser beam creator (left) and in the SFF (right).} \label{En_N01}
\end{figure}

\begin{figure}[ht!]
	\centering{
		\includegraphics[scale=0.5,trim={60 40 60 50},clip]{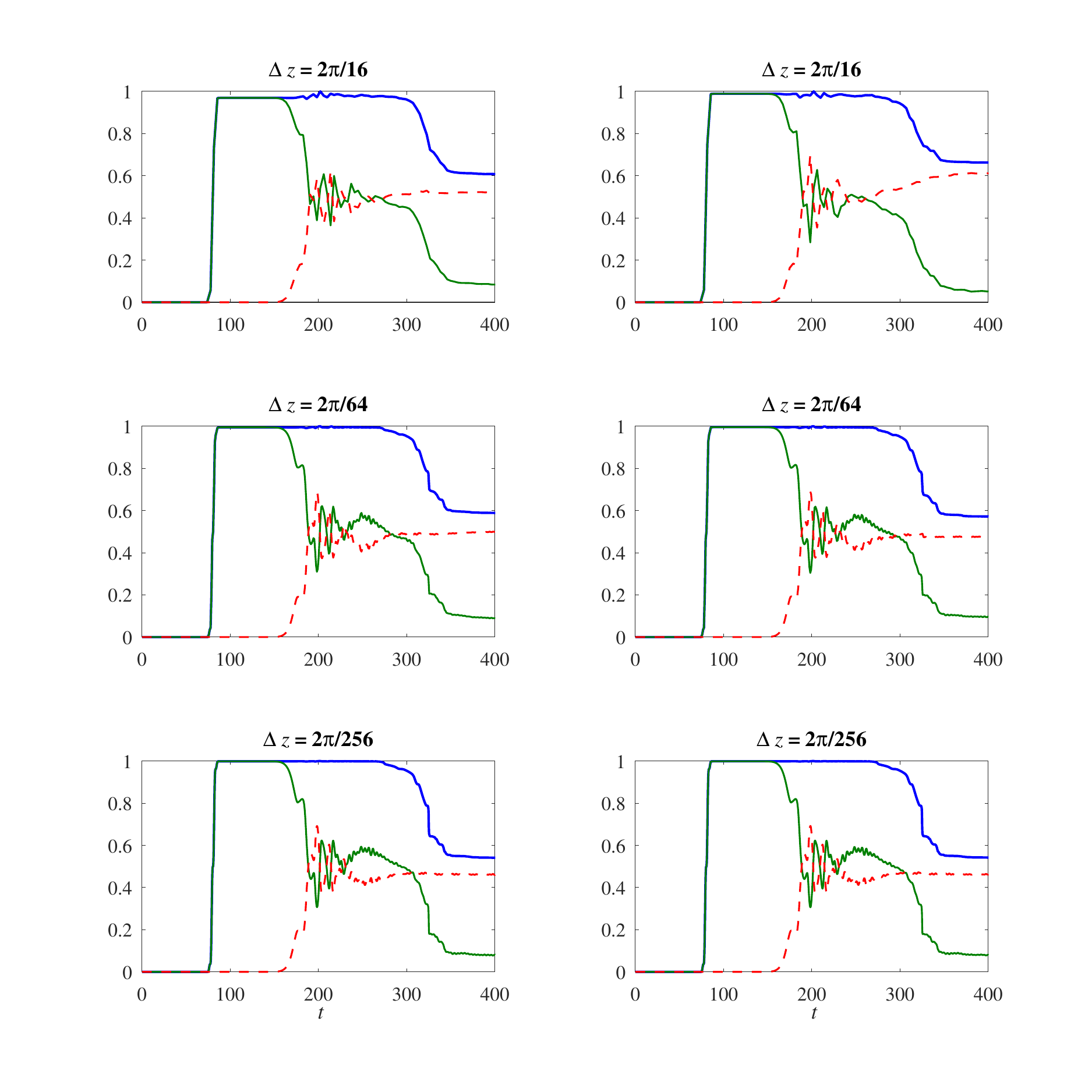}}
	\caption{The integrated kinetic energy of electrons (dashed red curves), the integrated energy of the electromagnetic field (thin green curves), the total energy (thick blue curves) in time, calculated for a pulse propagating in a plasma with a density of $n_{e0}=0.75$ in models with a laser beam creator (left) and in the SFF (right).} \label{En_N075}
\end{figure}

As stated in the title of this article, our main goal is to study the acceleration of plasma particles, so we now demonstrate how reducing the cell size affects the electromagnetic field energy and the kinetic energies of the particles. These results, which make it possible to understand the energy transition dynamics and confirm numerical energy conservation, are shown in figures~\ref{En_N01} and \ref{En_N075}. For convenience, the energies are normalized to the maximum value of the total energy. Note that the differences between these energy curves calculated within models (i) and (ii) in the case of $\Delta z = 2\pi/16$ are noticeable to the eye - for both plasma densities. This difference, however, disappears as the cell size decreases. To assess the convergence of energy solutions obtained in approaches (i) and (ii), we use the following norm
\begin{equation}\label{L2_t}
	\delta \mathcal{E}^{(i), (ii)}_{\Delta z}=\frac{1}{N}\sum_{n=1}^N|\mathcal{E}_{\Delta z}^{(i), (ii)}(t_n)-\mathcal{E}_{\frac{2\pi}{256}}^{(i), (ii)}(t_n)|,
\end{equation}
where $\mathcal{E}^{(i), (ii)}$ is either electromagnetic ($EM$) or kinetic ($kin$) energy integrated over the simulation domain. We also use the cross norm, calculated as 
\begin{equation}\label{L2_t_cross}
	\Delta \mathcal{E}=\frac{1}{N}\sum_{n=1}^N|\mathcal{E}_{\frac{2\pi}{256}}^{(i)}(t_n)-\mathcal{E}_{\frac{2\pi}{256}}^{(ii)}(t_n)|,
\end{equation}
to evaluate the convergence of the solutions of the different models as the cell size decreases.
The results, expressed as a percentage, are presented in the Table~\ref{tab:delta_E}.
\begin{table*}[h]
	\caption{Norms (in $\%$) estimating the convergence of 1D-solutions for the electromagnetic and kinetic energies according to formulas~\eqref{L2_t} and \eqref{L2_t_cross}.}
	\label{tab:delta_E}
	\begin{center}
\begin{tabular}{|c|c|c|c|c|c|c|c|c|c|c|}
	\hline
	&\multicolumn{4}{|c|}{$\Delta z = \frac{2\pi}{16}$} &  \multicolumn{4}{|c|}{$\Delta z = \frac{2\pi}{64}$}  &  \multicolumn{2}{|c|}{$\Delta z = \frac{2\pi}{256}$}  \\
	\hline
	&\multicolumn{2}{|c|}{$\delta \mathcal{E}_{EM}, \%$}& \multicolumn{2}{|c|}{$\delta \mathcal{E}_{kin}, \%$}  & \multicolumn{2}{|c|}{$\delta \mathcal{E}_{EM}, \%$} &  \multicolumn{2}{|c|}{$\delta \mathcal{E}_{kin}, \%$} & $\Delta \mathcal{E}_{EM}, \%$ & $\Delta \mathcal{E}_{kin}, \%$ \\
	\hline
	$n_{e0}$& $(i)$ &$(ii)$  &$(i)$  & $(ii)$ &$(i)$ &$(ii)$  &$(i)$  & $(ii)$ & &  \\
	\hline
	$0.1$&3.3  &2.9  &3.1  & 2.8 & 0.54 & 0.3 & 0.51 & 0.29 & 0.03 &0.03  \\
	\hline
	$0.75$&2.6  & 3.1 & 3.0 &4.7  &0.3  &0.9  &1.1  &0.6  &0.04  & 0.04 \\
	\hline
\end{tabular}
 \end{center}
\end{table*} 

Thus, even for a high plasma density, and even for the cell size $2\pi/64$, the averaged error does not exceed $1.1\%$. Concluding this section, we have shown that for the same 1D problem, the stability and convergence of PIC code solutions within the SFF are no worse than in the case of using the full-field approach.

\subsection{2D simulations}\label{2D}
In this section, as a reference solution, we use an analytical solution consisting of a Gaussian beam with a Gaussian time envelope.
We restrict ourselves to the 2D case in order to be able to perform convergence analysis for sufficiently small cell sizes, which in the 3D case becomes significantly more computationally expensive. Another fact that we must take into account is that even in a vacuum the electromagnetic field of a Gaussian beam is not purely transverse, so to compare solutions we need to add the $E_z$ component at least as a first order correction to $\epsilon =1/w_0k$, where $w_0$ is the waist parameter, as was done for the 3D case in \cite{Quesnel, Maltsev}. Thus, the solution in the paraxial approximation of Maxwell's equations for 2D Gaussian beam with focus at (0,0) in vacuum are
\begin{equation}\label{gauss2D}
	\begin{array}{rcl}
	E_x(x,z,t)&=&E_0\sqrt{\dfrac{w_0}{w(z)}}e^{-\frac{x^2}{w(z)^2}}\cos\Big(z-t+\dfrac{x^2}{2R(z)}-\eta(z)\Big)\cdot2^{-\frac{4(t-t_0-z)^2}{\text{FWHM}_T^2}},\\
	B_y(x,z,t) &=& E_x(x,z,t),\\
	E_z(x,z,t)&=&2E_0\frac{x}{w(z)}\sqrt{\dfrac{w_0}{w(z)}}e^{-\frac{x^2}{w(z)^2}}\sin\Big(z-t+\dfrac{x^2}{2R(z)}-\eta(z)\Big) \cdot 
    2^{-\frac{4(t-t_0-z)^2}{\text{FWHM}_T^2}},
	\end{array}
\end{equation}
where the spot size parameter
\begin{equation}
	w(z) = w_0\sqrt{1+\Big(\frac{z}{z_R}\Big)^2},
\end{equation}
the radius of wave front curvature
\begin{equation}
	R(z)=z\Big[1+\Big(\frac{z_R}{z}\Big)^2\Big],
\end{equation}
and the Gouy phase
\begin{equation}
	\eta(z)=\frac{1}{2}\arctan\Big(\frac{z}{z_R}\Big)
\end{equation}
are expressed through the dimensionless Rayleigh range $z_R=w_0^2/2$, where $w_0$ is divided by $L_r$, see Tab.~\ref{tab:units}.
To reduce the influence of non-paraxial effects in our simulations, we chose the ``relatively large'' beam waist parameter $w_0=16\lambda_c$, which leads to  $z_R= 512\pi^2$ in dimensionless units. The other beam parameters are FWHM$_T$=2$\pi$ and $E_0=4.86$. The simulation window size now is $2\pi\cdot[64, 32]$, with transversal size $L_x=128\pi$ large enough for such a wide $w_0$, however, the length in propagation direction $L_z=z_R/8\pi$ is twice as small as in the 1D simulations, and the simulation time $T_{\mathrm{sim}}=64\pi$. For convergence analysis we use the sequence of cell sizes $\Delta z = \dfrac{2\pi}{16\cdot2^m}$, $m=\{0,1,\dots,4\}$ with $\Delta x = \Delta z$, and $\Delta t = 0.99\Delta z/\sqrt{2}$ according to the CFL-condition.

As in Section~\ref{1D}, here we compare the solutions of two models: (i) the first is based on a laser beam creator, where at $z=0$ we set $B_y(x, 0, t)$ from \eqref {gauss2D} and $B_z(x, 0, t)=0$, taking into account that $z_{\mathrm{ghost}}=2\Delta z$; (ii) the second one uses SFF with prescribed electric and magnetic fields from \eqref{gauss2D}. Thus, for propagation in vacuum, the relative error for ``intensity'' defined as $I(x,z,t)=E_x^2(x,z,t)+E_z^2(x,z,t)$ at time points $t_n$ could be evaluated through the corresponding $L_1$-norms:
\begin{equation}
	\delta I(t_n) = \dfrac{\sum_{k,l}|I^{(i)}(x_k,z_l,t_n) -I^{(ii)}(x_k,z_l,t_n)|_{\Delta z}}{\sum_{k,l}I^{(ii)}(x_k,z_l,t_n)|_{\Delta z}}. 
\end{equation}
For a cell size $\Delta z = 2\pi/256$, it equals $0.7\%$ when passing through the focus point, and reaches the value of 1.2$\%$ by the end of the simulation time. We also compute errors by components using Frobenius norms
\begin{equation}
	\delta E(t_n) = \dfrac{\Big(\sum_{k,l}|E^{i}(x_k,z_l,t_n)-E^{ii}(x_k,z_l,t_n)|^2_{\Delta z} \Big)^{1/2}}{\Big(\sum_{k,l}|E^{ii}(x_k,z_l,t_n)|^2_{\Delta z} \Big)^{1/2}},
\end{equation}
where $E=E_{\{x,z\}}$. For transversal component $E_x(x,z)$ this norm slowly linearly increases again up to $1.2\%$ by the end of simulation time. As to longitudinal component $E_z(x,z)$, for $x=0$ we see good agreement between numerical solution (i) and purely analytical (ii) along the $z$-axis, where this component is 0. The dependence of $E_z$ on the transverse variable $x$ in the numerical solution agrees well with the analytical formula for the times $t=z$, however, after the envelope passes, we see that $E_z$ disappears more slowly than expected, probably due to reflection from $x_ {\mathrm{min}}$, $x_{\mathrm{max}}$ boundaries. However, since the longitudinal component is small, this discrepancy does not affect much the intensity calculation error, which, as noted above, does not exceed 1.2$\%$. Thus, the vacuum solutions for (i) and (ii) agree quite well.

Having proved a satisfactory matching of solutions in vacuum, we carried out simulations in plasma. In the $z$ direction, we use the same initial plasma profile as in 1D simulations, see Figure~\ref{super_gauss}, but with proportion shrunken twofold, while in the $x$-direction plasma is uniform. The maximum plasma density in the initial distribution is $n_{e0}=0.31$, and ionization processes are not taken into account in the test simulations. 

\begin{figure}[t!]
	\centering{
		\includegraphics[scale=0.65,trim={45 40 60 50},clip]{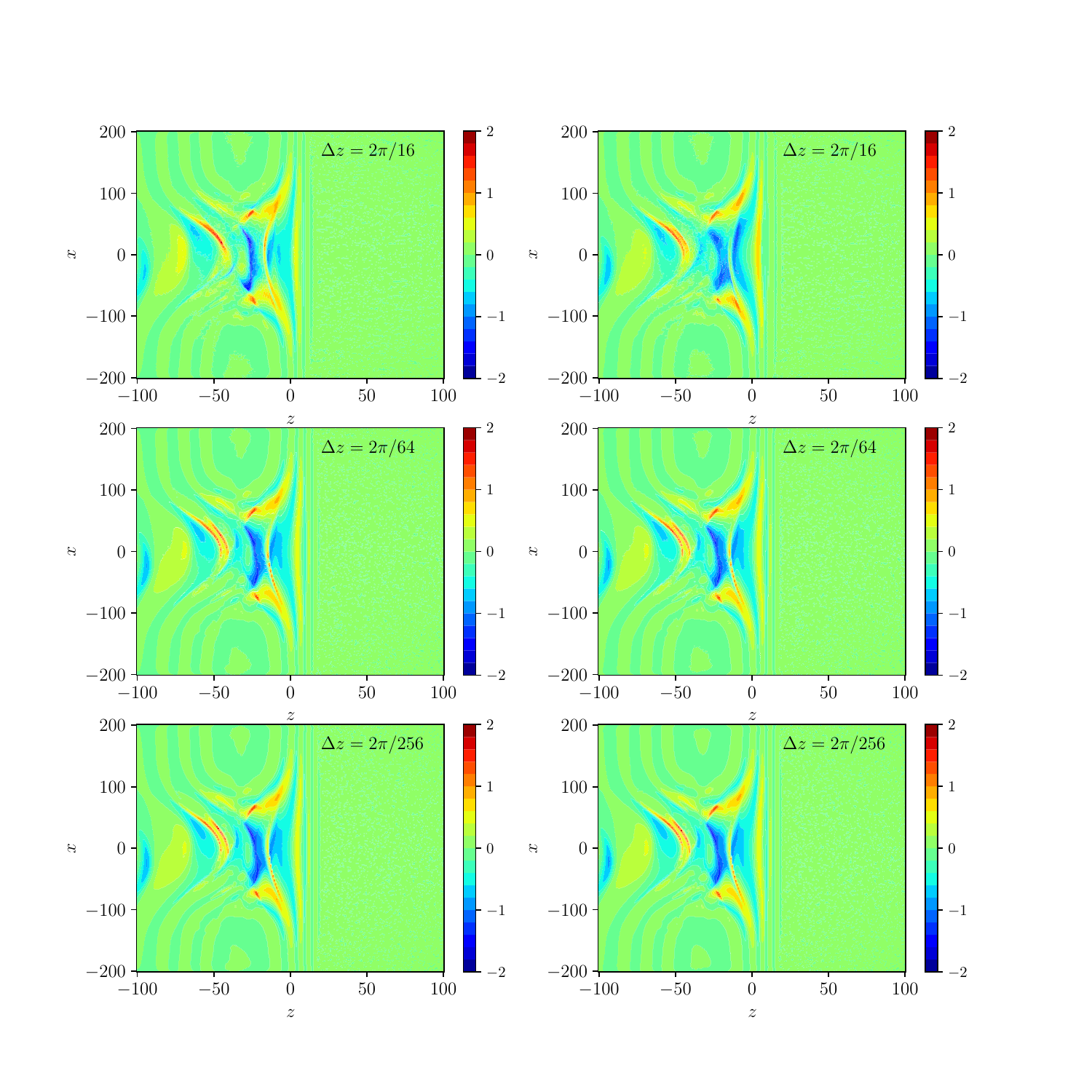}}
	\caption{Electric field components $E_x(x,z,t)$ calculated for different cell sizes $\Delta x = \Delta z = \{2\pi/16, 2\pi/64, 2\pi/256\}$ at time $t=145$ in the model (i) with laser creator (Left) and in SFF - model (ii) (Right) } \label{Ex_2D}
\end{figure}

\begin{figure}[t!]
	\centering{
		\includegraphics[scale=0.65,trim={45 40 60 50},clip]{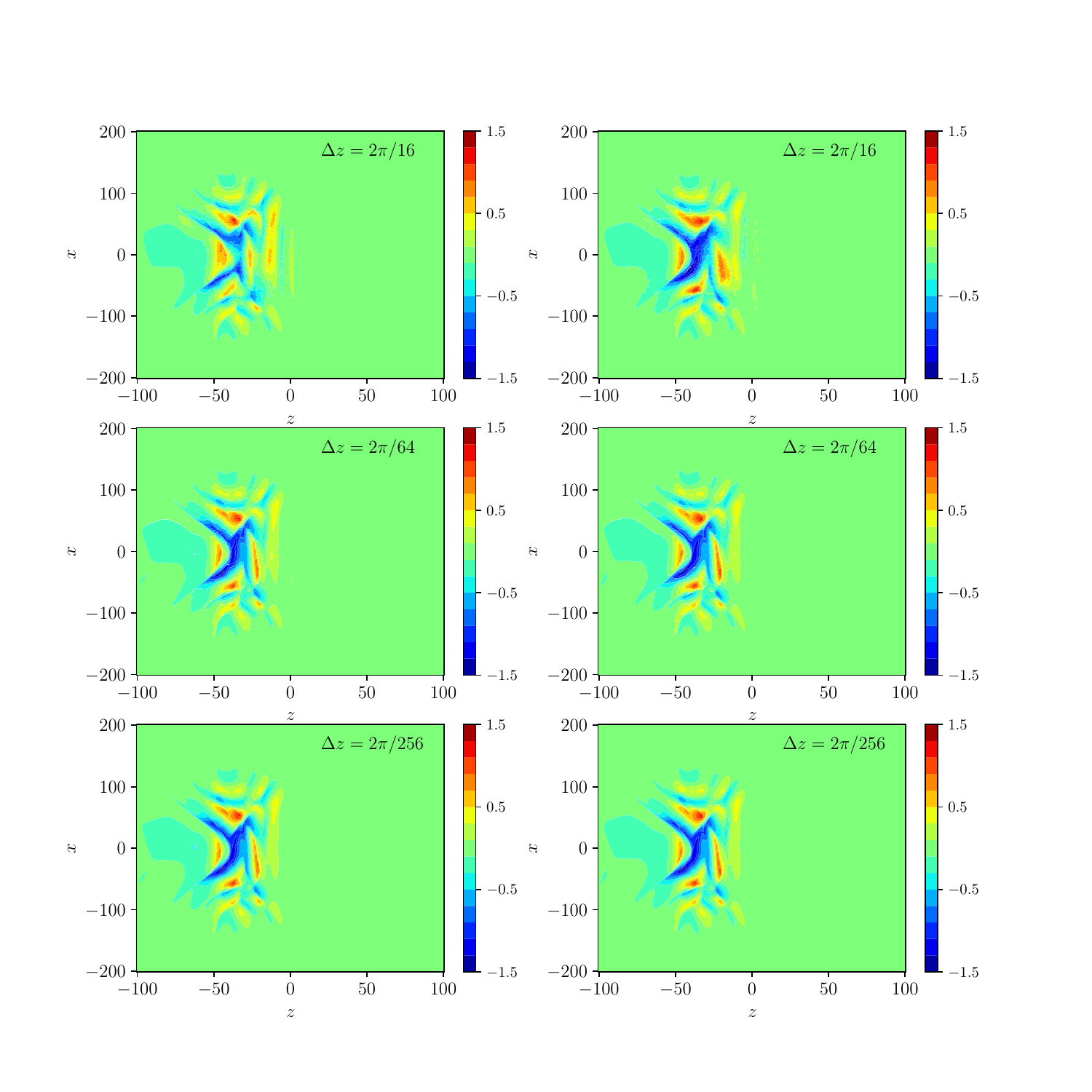}}
	\caption{Electric field components $E_z(x,z,t)$ calculated for different cell sizes $\Delta x = \Delta z =\{2\pi/16, 2\pi/64, 2\pi/256\}$ at time $t=145$ in the model (i) with laser creator (Left) and in SFF - model (ii) (Right)} \label{Ez_2D}
\end{figure}

\begin{figure}[ph!]
	\centering{
		\includegraphics[scale=0.5,trim={60 40 60 50},clip]{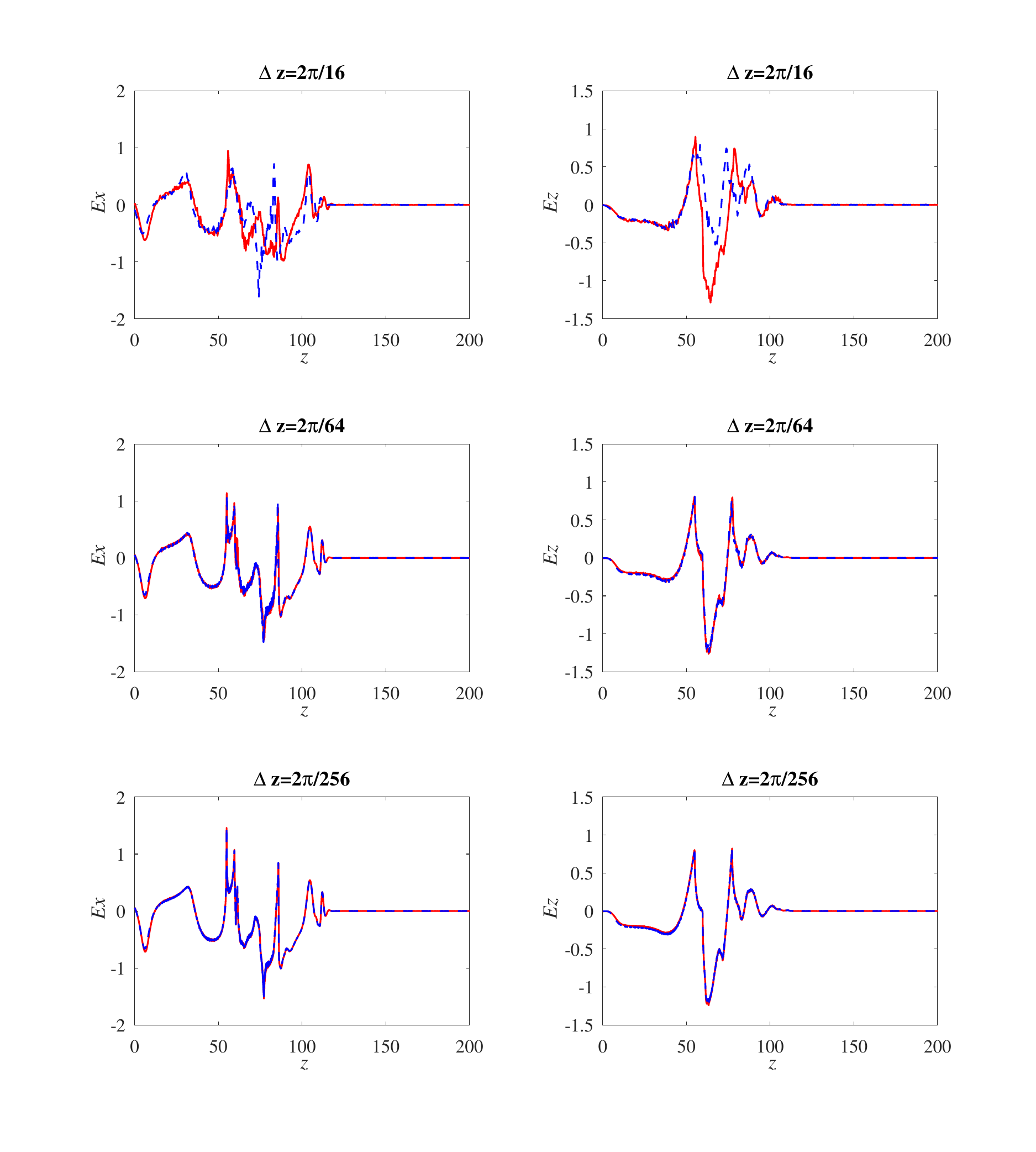}}
	\caption{Electric field components $E_x(0,z,t)$ and $E_z(0,z,t)$ at simulation time $t=145$ calculated for different cell sizes $\Delta z = \{2\pi/16, 2\pi/64, 2\pi/256\}$ for the problem of 2D Gaussian beam propagation in plasma with initial maximum density $n_{e0}=0.31$. Solutions shown in blue (dashed lines) are obtained for case (i) when laser field initially was set through the laser creator, and solutions in red correspond to the case (ii) when the incident field is prescribed.} \label{Esol_N031}
\end{figure}

We begin by demonstrating numerical solutions for the electric field components during the envelope propagation in plasma at time $t=145$ - shortly after it passes through the focal point in vacuum (corresponding to $t_f=133$), see Figures~\ref{Ex_2D} and \ref{Ez_2D}. We can observe convergence between solutions that become quite close to each other already at $\Delta z = 2\pi/64$. Moreover, if we look at these solutions in the $x=0$ section, Fig.~\ref{Esol_N031}, we can see that as the cell size decreases, both solutions continue to improve  (e.g. compare peaks in $E_x(0,z)$  for $\Delta z = 2\pi/64$ and $2\pi/256$). Also it seems that for relatively large cell size $\Delta z=2\pi/16$ the solution for $E_z$ computed in model (ii) is closer to both ``true'' solutions at $\Delta z = 2\pi/256$ than solution computed in model (i), i.e. when the laser beam creator is used.


\begin{figure}[t]
	\centering{
		\includegraphics[scale=0.5,trim={45 0 60 20},clip]{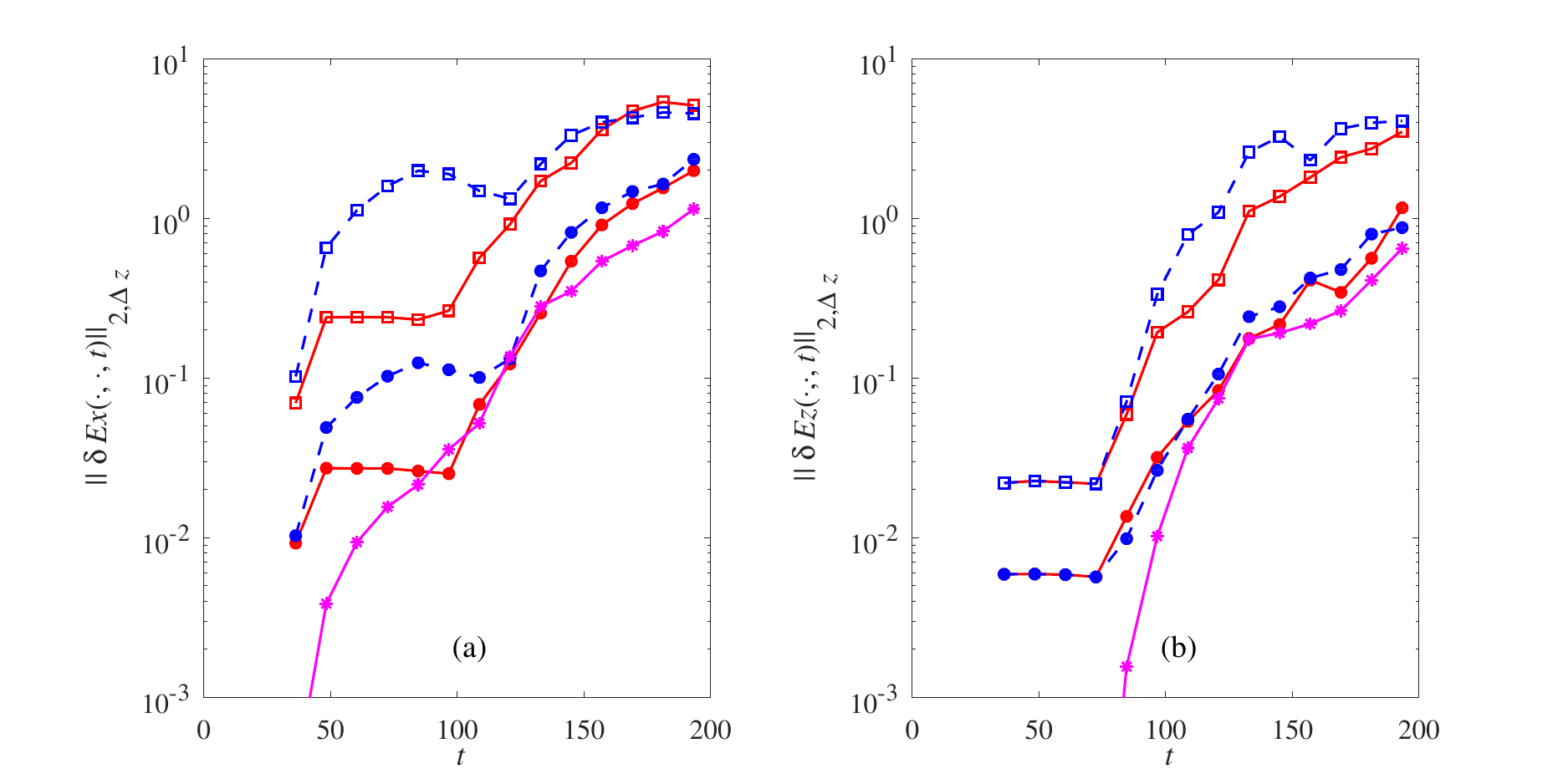}}
	\caption{Discrete $L_2$ norms for errors of solutions $E_x(0,z_i,t_n)$ - (a) and $E_z(0,z_i,t_n)$ - (b) computed according to \eqref{L2} for initial maximum plasma number density $n_{e0}=0.31$. Open squares are used for $\Delta z= 2\pi/16$, and filled circles - for $\Delta z= 2\pi/64$. The blue curves correspond to the case (i) - laser creator, the red curve corresponds to the applying of a prescribed field - case (ii). The norms, shown in magenta with asterisk markers, are calculated through the difference between the solutions for cases (i) and (ii) at $\Delta z=2\pi/256$.} \label{L2_2d}
\end{figure}

In order to demonstrate stability of solutions obtained in both models, in Fig.~\ref{L2_2d} we present $L_2(t_n)$ discrete norm for errors between numerical solutions $E_x(0,z_l,t_n)$ and $E_z(0,z_l,t_n)$ related to $\Delta z = 2\pi/16$ and $2\pi/64$ with respect to the ``exact'' solutions corresponding to $\Delta z = 2\pi/256$. Since, in the 2D case, both components were initially computed by solving Poisson's equation, the norms for solutions $E_x$ of model (ii) do not increase sharply from 0, like in the 1D case, although everything else is qualitatively similar to the norms shown in Fig.~\ref{L2_bothN}. Probably, the fact that in the 1D case we used a really exact solution, while in the 2D case the prescribed field is just an approximated solution with first order paraxial corrections, leads to some intersections between the norms for $\Delta z = 2\pi/64$ and $2\pi/ $256 in Fig.~\ref{L2_2d} (a), although it retains a consistent picture. As in the 1D case, with the help of discrete norms one can estimate the orders of convergence, which for all the following solutions $E_x(0,z)$, $E_x(x,0)$, $E_z(0,z)$ and $ E_z( x,0)$ are mostly around 1, sometimes deviating to 0.5 or 2.5.  
	
\begin{figure}[p]
	\centering{
		\includegraphics[scale=0.5,trim={60 40 60 50},clip]{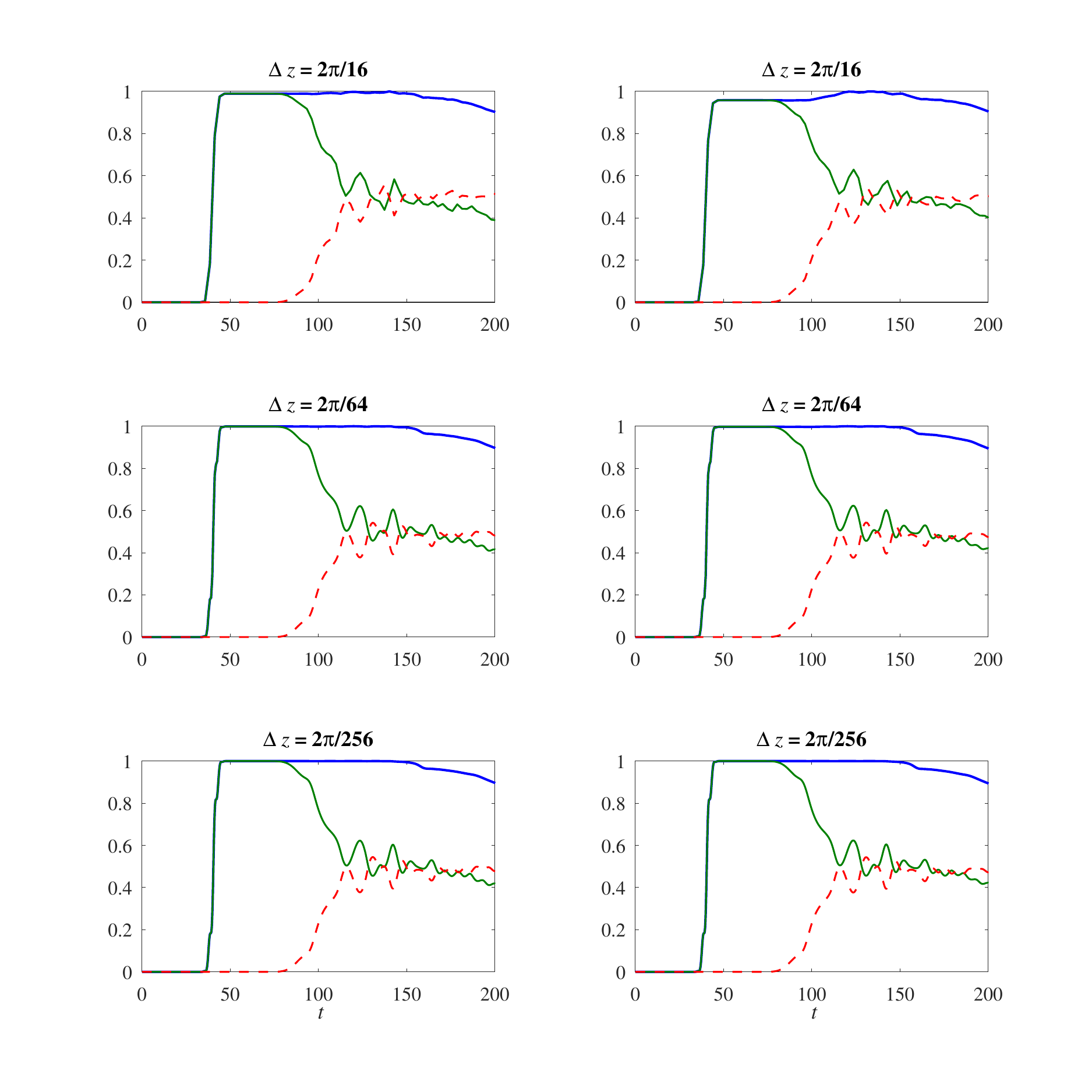}}
	\caption{The integrated kinetic energy of electrons (dashed red curves), the integrated energy of the electromagnetic field (thin green curves), the total energy (thick blue curves) in time, calculated for a 2D Gaussian beam propagating in a plasma with initial maximum density of $n_{e0}=0.31$ in models with a laser creator (left) and in the SFF (right).} \label{En_N031}
\end{figure}

Finally, we compare the dynamics between electromagnetic energy and electrons kinetic energy at different cell sizes, see Fig.~\ref{En_N031}. The first thing we can observe is that at $\Delta z = 2\pi/16$ the numerical energy balance at the stage of intense energy exchange between fields and electrons is disrupted, mainly for model (ii) - by $6 \%$, but, although to a lesser extent, for model (i) - by $2\%$. Fortunately, this numerical artifact can be successfully eliminated by reducing the cell size, as we observe in the same Fig.~\ref{En_N031} for the energy balance computed at $\Delta z= 2\pi/64$ and smaller. Further we use the same measures \eqref{L2_t} and \eqref{L2_t_cross} as in the 1D case to evaluate errors and convergence between the models, see Tab.~\ref{tab:delta_E_2D}. It seems that the cell size $\Delta z = 2\pi/64$ is already small enough to adequately describe particle dynamics in any of the models.

\begin{table*}[b]
	\caption{Norms (in $\%$) estimating the convergence of 2D-solutions for the electromagnetic and kinetic energies according to formulas~\eqref{L2_t} and \eqref{L2_t_cross}.}
	\label{tab:delta_E_2D}
	\begin{center}
		\begin{tabular}{|c|c|c|c|c|c|}
			\hline
			 \multicolumn{4}{|c|}{$\Delta z = \frac{2\pi}{64}$}  &  \multicolumn{2}{|c|}{$\Delta z = \frac{2\pi}{256}$}  \\			
			\hline
			$\delta \mathcal{E}_{kin}^{(i)}, \%$ &$\delta \mathcal{E}_{kin}^{(ii)}, \%$  &$\delta \mathcal{E}_{EM}^{(i)}, \%$ &$\delta \mathcal{E}_{EM}^{(ii)}, \%$&$\Delta \mathcal{E}_{EM}, \%$ &$\Delta \mathcal{E}_{kin}\%$  \\
			\hline
			 0.12 & 0.07 & 0.14 & 0.13 & 0.13 &0.15  \\			
			\hline
		\end{tabular}
	\end{center}
\end{table*} 

Again, these 2D results demonstrate the validity of the SFF method for inserting a laser beam into a simulation box, even at higher plasma densities. We now apply this method to a physically relevant configuration.

\newpage
\section{3D simulations for electrons acceleration in tightly-focused fields}\label{3D}
Here we demonstrate how the SFF works when simulating particle acceleration in tightly focused ($w_0 < \lambda_c$) ultrashort pulses. The procedure for obtaining exact analytical solutions for our modeling was proposed by April in \cite{April_2010}. While analytical expressions for radially polarized (RP) pulses were presented by April in the same work and even used to describe the direct acceleration of electrons in \cite{Marceau:12}, an analytical expression for linearly polarized (LP) pulse is now derived for this article in Section \ref{exact_LP}.

All simulations shown in this section were done at conditions similar to the experiment described in \cite{vallieres2022}, where the central wavelength $\lambda_c=1.8$ $\mu$m and we considered pulses with 2 cycles, see parameter $s$ in Section~\ref{exact_LP}. The simulation box size is $2\pi\cdot[24, 24, 32]$ in plasma units, assuming the pulse propagates in the $z$-direction. The plasma profile in the $z$-direction is the same as for 2D-simulations with uniformity in the $x$- and $y$-directions. Under standard conditions, the number density of the neutral molecular nitrogen is estimated as $2.687\cdot10^{25}$ m$^{-3}$. We assume that a relatively low-intensity leading edge of the laser pulse dissociates molecules and ionizes them at least twice \cite{vallieres2022}, so we start our simulations with $n_{e0}=0.31$ in units of critical density $N_r$. Since here we are simulating a real experiment, we also take into account tunneling ionization by the field using the corresponding Smilei module \cite{10.1063/1.3559494}.

We set the same spatial steps for all directions, so that now $\Delta t = 0.99\Delta z/\sqrt{3}$ to safely satisfy the CFL-condition. Certainly, in 3D case we cannot perform so many tests as in our 1D and even 2D simulations, however we still compare results for cell size $\Delta=2\pi/16$ and $2\pi/32$. Since we cannot  set a tightly focused pulse using a laser beam creator (method (i) throughout Section~\ref{ConvAn}), here we only compare the field solutions obtained in SFF for these two cell sizes.

\subsection{Exact solution for ultrashort linearly polarized tightly-focused pulse in vacuum}\label{exact_LP}
We follow the strategy presented by A.April in his work \cite{April_2010} for closed-form solutions for the type of pulsed beam named in the title of this section. Thus, the field components of an isodiffracting pulse propagating in the $z$-direction are
\begin{equation}\label{components}
	\begin{array}{lcl}		
		E_x=\partial^2_{xx}\Psi-c^{-2}\partial^2_{tt}\Psi+c^{-1}\partial^2_{tz}\Psi, && cB_x = \partial^2_{xy}\Psi, \\
		E_y = \partial^2_{xy}\Psi, 
		&&cB_y=\partial^2_{yy}\Psi-c^{-2}\partial^2_{tt}\Psi+c^{-1}\partial^2_{tz}\Psi, \\
		E_z=\partial^2_{zx}\Psi-c^{-1}\partial^2_{tx}\Psi,&& 
		cB_z=\partial^2_{zy}\Psi-c^{-1}\partial^2_{ty}\Psi,
	\end{array}
\end{equation}
with phasor 
\begin{equation}\label{phasor}
	\Psi(x,y,z,t) = \dfrac{\Psi_0}{\widetilde{R}}[f(\tilde{t}_+)-f(\tilde{t}_-)],
\end{equation}
expressed for Poisson-like spectrum with central frequency $\omega_0=kc$ and constant phase $\phi_0$ using the function
\begin{equation}\label{poisson}
	f(t)= e^{i\phi_0}\Big(1-\dfrac{i\omega_0t}{s}\Big)^{-(s+1)},
\end{equation}
whose arguments
\begin{equation}
	\tilde{t}_\pm = t \pm \widetilde{R}/c+ia/c,
\end{equation}
and the complex spherical radius
\begin{equation}	
	\widetilde{R}=\sqrt{x^2+y^2+(z+ia)^2}.
\end{equation}
In the expressions above, $\Psi_0$, $a$, $s$ are the real parameters of the pulse associated with its amplitude, focus spot size, and envelope duration, respectively, - same as in case of RP pulse, see \cite{Marceau:12}. This means that there is no need to apply the envelope separately, it is already built into the analytical solution. For example, $s=10$ results in 2 optical cycles, and with $s=60$ the envelope includes 6 cycles. Knowing the parameter $a$, we can express the size of the beam waist in terms of the central wavelength, since $k^2w_0^2=2(\sqrt{1+(ka)^2}-1)$. Finally, the amplitude can be expressed in terms of the peak power of the pulse $P_{\mathrm{peak}}$, wavenumber $k$ and two other parameters $a$ and $s$, see below.

Substituting \eqref{phasor} to \eqref{components} and passing to dimensionless variables and functions we obtain
\begin{eqnarray}
	\nonumber	E_x &=&\dfrac{x^2}{r^2}\dfrac{\Psi_0}{a^3} \Re\Big\{ \dfrac{\sin^2\tilde{\theta}}{\widetilde{\mathcal{R}}}\Big(\dfrac{3\mathcal{G}^{(0)}_-}{\widetilde{\mathcal{R}}^2}-\dfrac{3\mathcal{G}^{(1)}_+}{\widetilde{\mathcal{R}}}+ \mathcal{G}^{(2)}_- \Big)\Big\}\\
	&-&
	\dfrac{\Psi_0}{a^3} \Re\Big\{\dfrac{1}{\widetilde{\mathcal{R}}}\Big(\dfrac{\mathcal{G}^{(0)}_-}{\widetilde{\mathcal{R}}^2}+\dfrac{\mathcal{G}^{(1)}_-\cos\tilde{\theta}}{\widetilde{\mathcal{R}}} -\dfrac{\mathcal{G}^{(1)}_+}{\widetilde{\mathcal{R}}}+\mathcal{G}^{(2)}_--\mathcal{G}^{(2)}_+\cos\tilde{\theta}\Big)\Big\},\\
	\nonumber \\
	E_y &=& \dfrac{xy}{r^2}\dfrac{\Psi_0}{a^3}\Re\Big\{\dfrac{\sin^2\tilde{\theta}}{\widetilde{\mathcal{R}}}\Big(\dfrac{3\mathcal{G}^{(0)}_-}{\widetilde{\mathcal{R}}^2}-\dfrac{3\mathcal{G}^{(1)}_+}{\widetilde{\mathcal{R}}}+ \mathcal{G}^{(2)}_- \Big)\Big\},\\
	\nonumber \\
	\nonumber E_z &=&\dfrac{x}{r}\dfrac{\Psi_0}{a^3}\Re\Big\{ \dfrac{\sin\tilde{\theta}}{\widetilde{\mathcal{R}}}\Big(\dfrac{3\mathcal{G}^{(0)}_-\cos\tilde{\theta}}{\widetilde{\mathcal{R}}^2}+\dfrac{\mathcal{G}^{(1)}_-}{\widetilde{\mathcal{R}}}-\dfrac{3\mathcal{G}^{(1)}_+\cos\tilde{\theta}}{\widetilde{\mathcal{R}}}+ \mathcal{G}^{(2)}_-\cos\tilde{\theta}- \mathcal{G}^{(2)}_+ \Big)\Big\},\\
\end{eqnarray}
\begin{eqnarray}  
	cB_x &=& \dfrac{xy}{r^2}\dfrac{\Psi_0}{a^3}\Re\Big\{\dfrac{\sin^2\tilde{\theta}}{\widetilde{\mathcal{R}}}\Big(\dfrac{3\mathcal{G}^{(0)}_-}{\widetilde{\mathcal{R}}^2}-\dfrac{3\mathcal{G}^{(1)}_+}{\widetilde{\mathcal{R}}}+ \mathcal{G}^{(2)}_- \Big)\Big\},\\
	\nonumber \\
	\nonumber	cB_y &=&\dfrac{y^2}{r^2}\dfrac{\Psi_0}{a^3} \Re\Big\{ \dfrac{\sin^2\tilde{\theta}}{\widetilde{\mathcal{R}}}\Big(\dfrac{3\mathcal{G}^{(0)}_-}{\widetilde{\mathcal{R}}^2}-\dfrac{3\mathcal{G}^{(1)}_+}{\widetilde{\mathcal{R}}}+ \mathcal{G}^{(2)}_- \Big)\Big\}\\
	&-&
	\dfrac{\Psi_0}{a^3}\Re\Big\{\dfrac{1}{\widetilde{\mathcal{R}}}\Big(\dfrac{\mathcal{G}^{(0)}_-}{\widetilde{\mathcal{R}}^2}+\dfrac{\mathcal{G}^{(1)}_-\cos\tilde{\theta}}{\widetilde{\mathcal{R}}} -\dfrac{\mathcal{G}^{(1)}_+}{\widetilde{\mathcal{R}}}+\mathcal{G}^{(2)}_--\mathcal{G}^{(2)}_+\cos\tilde{\theta}\Big)\Big\},\\
	\nonumber\\
	\nonumber cB_z &=&\dfrac{y}{r}\dfrac{\Psi_0}{a^3}\Re\Big\{ \dfrac{\sin\tilde{\theta}}{\widetilde{\mathcal{R}}}\Big(\dfrac{3\mathcal{G}^{(0)}_-\cos\tilde{\theta}}{\widetilde{\mathcal{R}}^2}+\dfrac{\mathcal{G}^{(1)}_-}{\widetilde{\mathcal{R}}}-\dfrac{3\mathcal{G}^{(1)}_+\cos\tilde{\theta}}{\widetilde{\mathcal{R}}}+ \mathcal{G}^{(2)}_-\cos\tilde{\theta}- \mathcal{G}^{(2)}_+ \Big)\Big\},\\
\end{eqnarray}
where functions
\begin{equation}
	\mathcal{G}_\pm^{(n)}=g^{(n)}(\tilde{\tau}_+)\pm g^{(n)}(\tilde{\tau}_-),\quad n=0,1,2
\end{equation} 
are expressed through function ($n=0$) or derivatives ($n=1,2$)
\begin{equation}\label{poisson_dstr}
	g^{(n)}(\tau)=e^{i\phi_0}\dfrac{\Gamma(s+n+1)}{\Gamma(s+1)}\Big(\dfrac{ika}{s}\Big)^n\Big(1-\dfrac{i\tau}{s}\Big)^{-(s+n+1)},
\end{equation}
with arguments
\begin{equation}
	\tau_\pm = \tau + ka(\pm\widetilde{\mathcal{R}}+i),
\end{equation}
where $\tau = \omega_0t$, $\widetilde{\mathcal{R}} = \sqrt{\rho^2+(\zeta + i)^2}$, $\rho= r/a$, $\zeta = z/a$. Also in component expressions we use the notation: $\sin\tilde{\theta}=\dfrac{\rho}{\widetilde{\mathcal{R}}}$, $\cos\tilde{\theta} = \dfrac{\zeta + i}{\widetilde{\mathcal{R}}}$. 

Now, as was done in \cite{Marceau:12}, we estimate $\Psi_0$ in terms of averaged peak laser power. The time-averaged $z$-component of the Poynting vector is
\begin{equation*}
	\left\langle S_z(x,y) \right\rangle = \dfrac{1}{2\mu_0}\Re\{E_xB^*_y - E_yB^*_x\}.
\end{equation*}
For the focal plane, we set $\zeta=0$ and the focus passing time $\tau=0$, then
\begin{eqnarray}
	\nonumber	P_{\mathrm{peak}}&=&\frac{\pi \Psi_0^2}{\mu_0a^4c}\int_0^\infty\dfrac{1}{|\mathcal{R}_0|^2}\Big[\Re\Big\{\Big(\dfrac{3\mathcal{G}^{(0)}_-}{\mathcal{R}^2_0}-\dfrac{3\mathcal{G}^{(1)}_+}{\mathcal{R}_0}+ \mathcal{G}^{(2)}_- \Big)\Big(\cos\tilde{\theta} \Big(\dfrac{\mathcal{G}^{(1)}_-}{\mathcal{R}_0} - \mathcal{G}^{(2)}_+\Big)^*\\
	\nonumber &-&\Big(\dfrac{\mathcal{G}^{(0)}_-}{\mathcal{R}^2_0}-\dfrac{\mathcal{G}^{(1)}_+}{\mathcal{R}_0}+\mathcal{G}^{(2)}_-\Big)^*\Big\}|\sin\tilde{\theta}|^2\\
	\nonumber
	&+&\Big|\cos\tilde{\theta} \Big(\dfrac{\mathcal{G}^{(1)}_-}{\mathcal{R}_0} - \mathcal{G}^{(2)}_+ \Big)+ \Big(\dfrac{\mathcal{G}^{(0)}_-}{\mathcal{R}^2_0}-\dfrac{\mathcal{G}^{(1)}_+}{\mathcal{R}_0}+\mathcal{G}^{(2)}_-\Big)  \Big|^2\Big]\rho d\rho\equiv\frac{\pi \Psi_0^2}{\mu_0a^4c}\mathcal{I}(s,ka).\\
\end{eqnarray}
Hence the amplitude is
\begin{equation}
	\Psi_0=\sqrt{\dfrac{\mu_0ca^4P_{\mathrm{peak}}}{\pi\mathcal{I}(s,ka)}}.
\end{equation}

\subsection{Acceleration by linearly polarized pulse}
\begin{figure}[ph!]
	\centering{
		\includegraphics[scale=0.60,trim={40 40 40 65},clip]{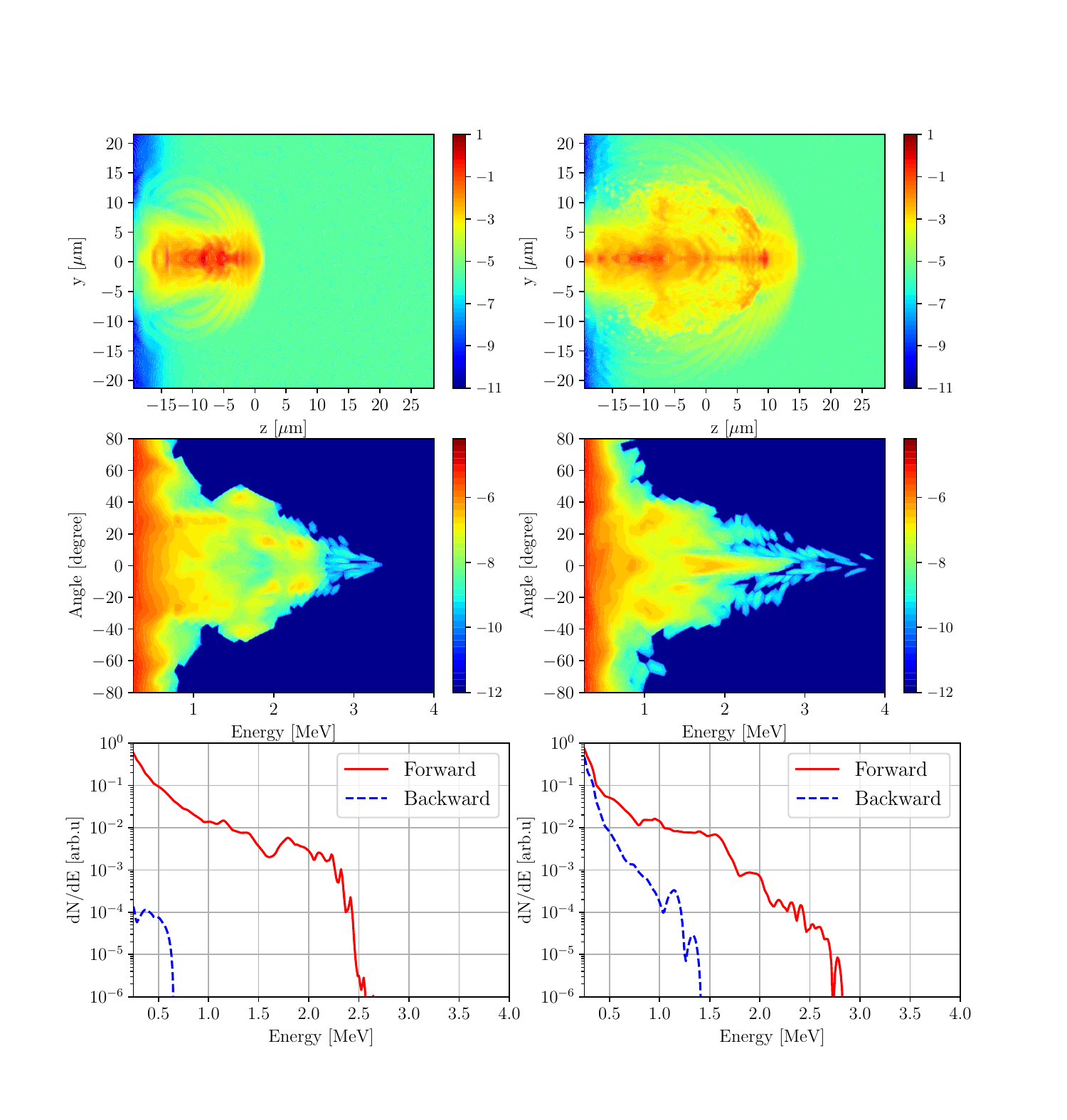}}
	\caption{Simulation of electron acceleration by LP pulse with $P_{\mathrm{peak}}=2\cdot10^{11}$ W and focal spot size of $w_\text{FWHM} = 2$ $\mu$m, on grid $\Delta z = 2\pi/32$. Results correspond to two time points: left column,  (a), (c), (e), is for  $t_1=132$ fs and right one (b), (d), (f) - for $t_2=180$ fs. Insets (a) and (b) show the electron kinetic energy distribution, (c) and (d) show the electron energy-angle distribution for forward propagation, (e) and (f) show the energy distribution for the angle $20^\circ$ forward (solid red line) and backward (dashed red line) wrt the pulse propagation.}\label{LP32}
\end{figure}

The geometry of the numerical experiment is as follows. The polarization axis is the $x$ axis, and we can see the distribution of the kinetic energy of electrons in the $(y,z)$ plane at some preselected times. In the same plane, we bin the particles to get their energy and angle distribution, where the angle is measured from the direction $z$. In particular, we are interested in the electron energy distribution corresponding to the angle $20^{\circ}$, at which the detection is performed in the experiment \cite{vallieres2022}. We present these results in Fig.~\ref{LP32} for $t_1=132$ fs, which corresponds to the time it takes for the envelope to pass through the focus in a vacuum, and for some later time $t_2=180$ fs. When propagating in a vacuum at a peak power of $P_{\mathrm{peak}}=2\cdot10^{11}$ W, the maximum intensity at the focal position is equal to $4.8\cdot10^{18}$ W/cm$ ^2 $, while in plasma it can reach $10.6\cdot10^{18}$ W/cm$^2$ due to field amplification by plasma currents generated by the laser. Moreover, this occurs not only due to the strengthening of the field component $E_x$, but also due to the strengthening of the $E_z$ component to somewhat smaller, but comparable values.  
At this intensity, the maximum kinetic energy of electrons is determined not only by acceleration due to the ponderomotive force, although the predominant acceleration in the direction of propagation of the pulse, rather than backward, indicates a significant contribution of this mechanism.
Another piece of evidence for the contribution of the ponderomotive force is electron density bunches, which can be seen in insets (a) and (b), \cite{vallieres2022}. However, even after passing the focus, the fields generated by the plasma currents have significant amplitudes and continue to accelerate the electrons. This may be exploited further to reach higher electron energies or flux.

\begin{figure}[t]
	\centering{
		\includegraphics[scale=0.7,trim={0 0 0 0},clip]{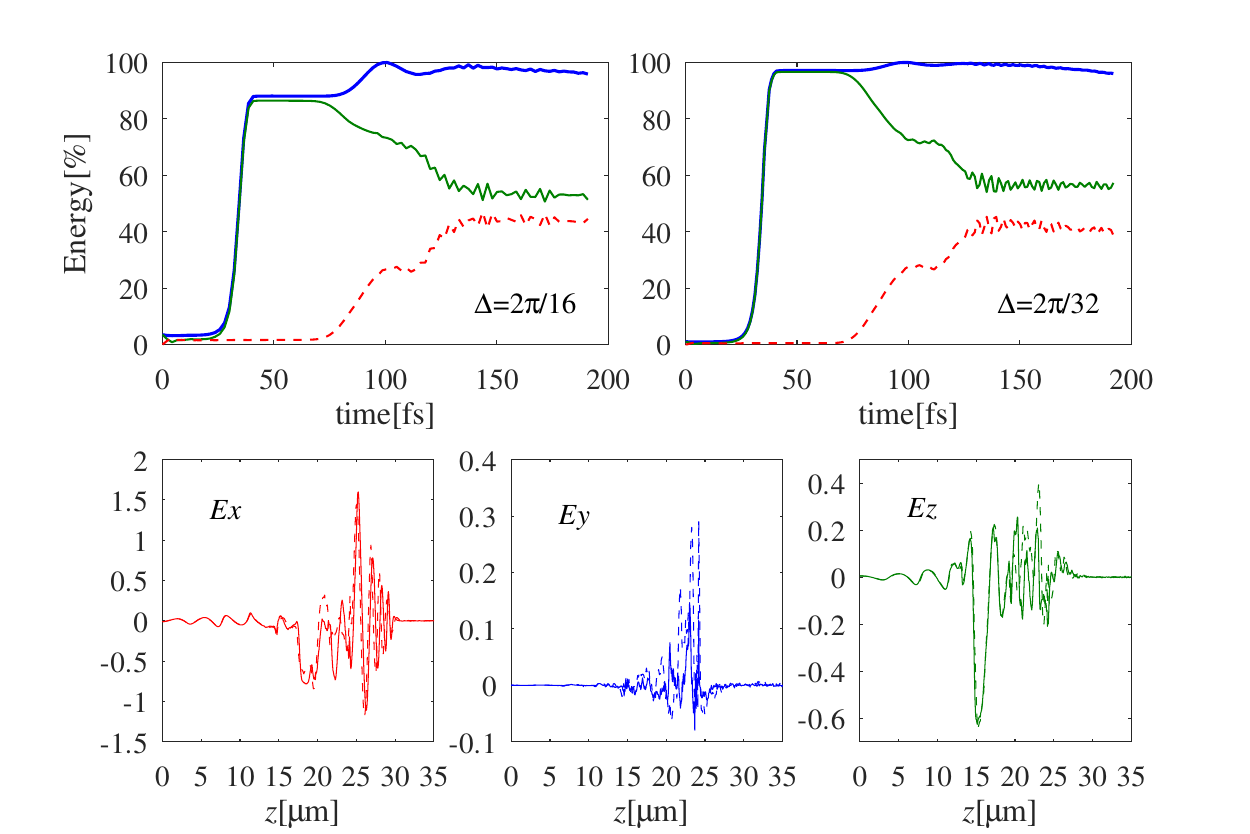}}
	\caption{Upper row: integrated total (thick blue line), electromagnetic (thin green line) and electrons kinetic (dashed red line) energies computed for cell sizes $\Delta=2\pi/16$ (left) and $\Delta=2\pi/32$ (right). All other parameters are the same as in Fig.~\ref{LP32}. Low row: electric field components at time $t_1=132$ fs for $\Delta z = 2\pi/16$ (dashed lines) and $\Delta z = 2\pi/32$ (solid lines) }\label{LP32_conv}
\end{figure}

To evaluate the convergence of solutions as the cell size decreases, Fig.~\ref{LP32_conv} shows the dynamics of various types of integrated energies over time and numerical solutions for the electric field components at time $t_1$. From the energy curves, we can conclude that the solution is stable and its accuracy increases with decreasing $\Delta z$, improving the conservation of the total numerical energy. The insets in the bottom row show that the electric field components for cell sizes $\Delta z = 2\pi/16$ and $2\pi/32$ have much in common, and decreasing the cell size corrects the field amplitudes, in particular, locally reducing them. Certainly, the results would be more accurate if we could run simulations for $\Delta z = 2\pi/64$, but even for $\Delta z = 2\pi/32$ they already take 18 hours on 5120 cores. For the same reason, we, unfortunately, cannot conduct a proper analysis of convergence on the required set of solutions.

\subsection{Acceleration by radially polarized pulse}
\begin{figure}[ph!]
	\centering{
		\includegraphics[scale=0.60,trim={40 20 40 65},clip]{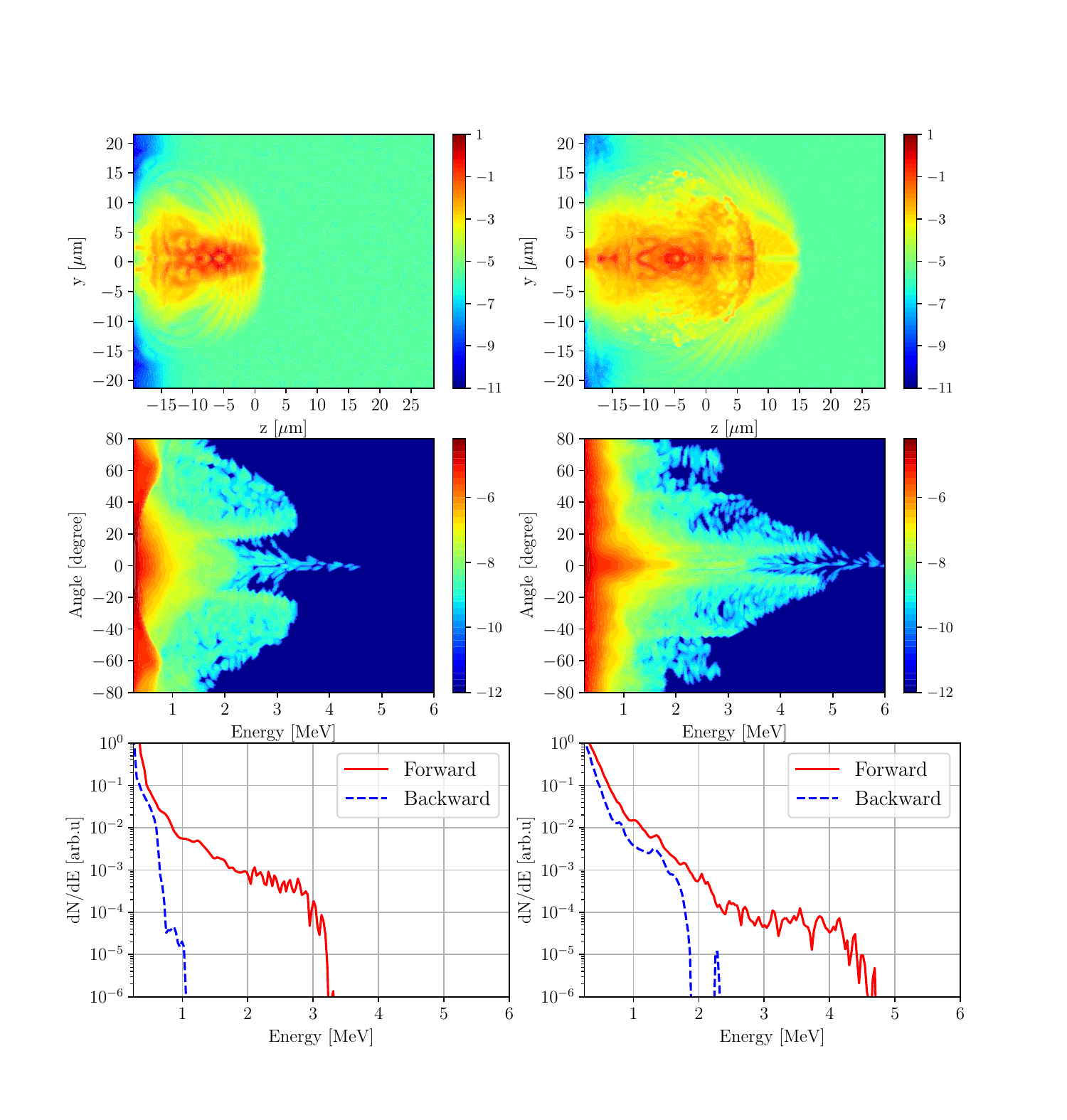}}
	\caption{Simulation of electron acceleration by RP pulse with $P_{\mathrm{peak}}=4\cdot10^{11}$ W and focal spot size of $w_{FWHM} = 2$ $\mu$m, $\Delta z = 2\pi/32$. Results correspond to two time points: left column,  (a), (c), (e), is for  $t_1=132$ fs and right one (b), (d), (f) - for $t_2=180$ fs. Insets (a) and (b) show the electron kinetic energy distribution, (c) and (d) show the electron energy-angle distribution  for forward propagation, (e) and (f) show the energy distribution for the angle $20^\circ$ forward (solid red line) and backward (dashed red line) wrt the pulse propagation.}\label{RP32}
\end{figure}

Here we use the same geometry for the numerical experiment, although the choice of the $(x, y)$ plane for the RP pulse is arbitrary, since the solution in plasma preserves cylindrical symmetry. We show the results of simulations on grid with $\Delta z = 2\pi/32$ in Fig.~\ref{RP32}. With the higher peak power chosen for this simulation, $P_{\mathrm{peak}}=4\cdot10^{18}$ W in vacuum at the focus, we see lower field amplitudes than for a LP pulse, now $2.7 \cdot10^{18}$ W/cm$^2$. However, due to the steeper gradient in the case of the RP pulse \cite{Salamin:06}, we see more electrons involved in the acceleration process and higher maximum energies. As in the case of interaction with the LP pulse, we see signs of the ponderomotive force action, but other acceleration mechanisms must necessarily be involved. Note that in plasma the amplitude of $E_z$-component could even exceed $E_x$ and $E_y$. This happens due to the larger longitudinal fields in this configuration, which yields stronger ponderomotive force \cite{Salamin:06}.

\begin{figure}[t]
	\centering{
		\includegraphics[scale=0.7,trim={0 0 0 0},clip]{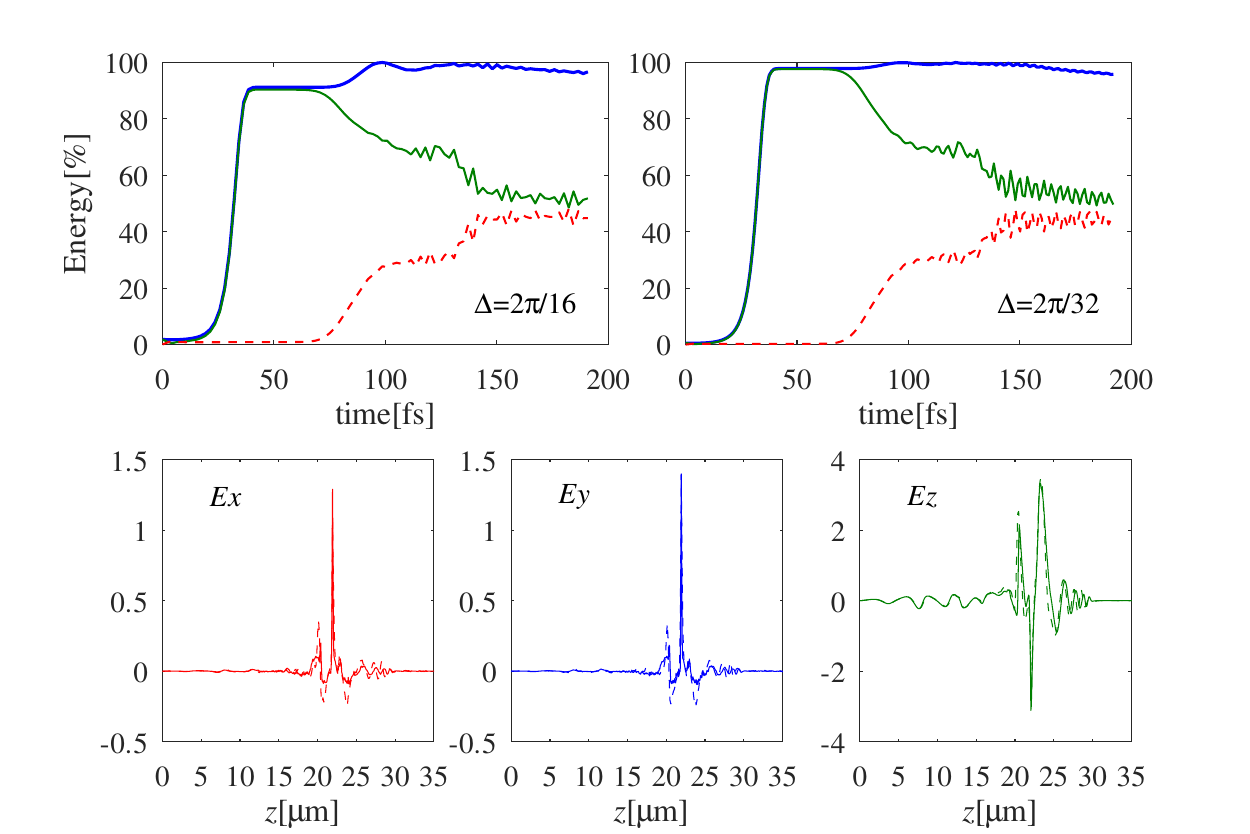}}
	\caption{Upper row: integrated total (thick blue line), electromagnetic (thin green line) and electrons kinetic (dashed red line) energies computed for cell sizes $\Delta=2\pi/16$ (left) and $\Delta=2\pi/32$ (right). All other parameters are the same as in Fig.~\ref{RP32}. Low row: electric field components at time $t_1=132$ fs for $\Delta z = 2\pi/16$ (dashed lines) and $\Delta z = 2\pi/32$ (solid lines) }\label{RP32_conv}
\end{figure}

Finally we compare the solutions for energy dynamics and field components obtained for different cell sizes in Fig.~\ref{RP32_conv}. We can draw the same conclusion as for the LP simulation: a decrease in $\Delta z$ improves the accuracy of energy conservation and calculation of field components.

\section{Conclusion}\label{concl}
In this article we provided arguments in favor of using the SFF with PIC codes and gave examples of solutions. This development of the model allows the implementation of non-paraxial laser beam configurations, such as tightly-focused fields or other complex field configurations (like the ones in \cite{Vallieres:23}), which are in great demand today for simulating experimental conditions.
In addition, the implementation of this method is relatively simple, given that most PIC codes already offer the possibility to add prescribed fields with arbitrary spatio-temporal dependences. 

Numerical analysis of 1D and 2D problems for Gaussian envelope and Gaussian beam, which can be solved not only in the pure PIC approach, but also using the PIC approach with SFF, showed that a decrease in the cell size leads to the convergence of the solutions obtained in both models. Finally, within the PIC-SFF approach we considered a problem of electron acceleration in linearly and radially polarized tightly-focused laser pulses. Again, the solutions looks stable and their accuracy increase with decreasing cell size.

In terms of computational performance, the computation times of both models in 1D was the same. However, for 2D simulations, when the complexity of the solution is higher and when parallelization is required, we observed a difference: the task in SFF is about 1.5 times slower than for the PIC code itself. This may be explained by the parallellization technique implemented in Smilei, which is based on a hybrid OpenMP-MPI framework. Our implementations of analytical models were not parallellized with OpenMP, which could lead to many threads reading the same memory, potentially deteriorating the performance. 
Future research will focus on finding a solution to remedy this problem. In 3D modeling, we even see a performance difference between LP and RP pulses, since the solution in the former case has more components and requires more computation operations. 

Despite the overhead in calculating the analytical solution numerically, we believe that the SFF-PIC method will be useful for a number of applications where realistic laser field configurations are required. This is especially useful for experiments in high intensity laser-matter interactions, where the laser field is focused tightly. Therefore, in the future, we plan to use this technique to describe high-intensity laser-plasma interactions and optimize electron acceleration.

\section*{Acknowledgments}
Authors thank Shanny Pelchat-Voyer for providing C-code to compute the exact solution for RP pulse in vacuum. They also thank Emmanuel d'Humi\`{e}res for stimulating discussions and encouragements, and Emmanuel Lorin for insightful comments on the general approach. This research was enabled in part by support provided by the Digital Research Alliance of Canada  \url{alliancecan.ca.}

\clearpage

\bibliography{mybibfile}

\end{document}